\documentclass[a4paper,UKenglish]{lipics-2016}
\usepackage[author=anonymous,nomargin,marginclue,footnote,status=final]{fixme}
\FXRegisterAuthor{ls}{als}{LS}
\FXRegisterAuthor{dp}{atm}{dp}
\FXRegisterAuthor{ud}{aud}{ud}
\usepackage{amsmath}
\usepackage{amsthm}
\usepackage{MnSymbol}
\usepackage{stmaryrd}
\usepackage{bussproofs}
\usepackage[all]{xy}
\usepackage[usenames]{xcolor}
\usepackage{amsfonts}
\usepackage[utf8]{inputenc}

\usepackage{tikz-cd}

\newcommand{\can}{\mathsf{can}}
\newcommand{\Ats}{\mathcal{S}}
\newcommand{\impl}{\Rightarrow}
\newcommand{\myparagraph}[1]{\par\medskip\noindent\textbf{\sffamily #1}\quad}

\newcommand{\Wfun}[1]{\mathsf{S}_{#1}}

\newcommand{\hgt}{\mathsf{ht}}
\newcommand{\Th}{\mathsf{Th}}
\newcommand{\rank}{\mathsf{rk}}
\newcommand{\range}{\mathsf{rg}}
\newcommand{\domain}{\mathsf{dom}}
\newcommand{\BA}{\mathfrak{A}}
\newcommand{\Lang}{\mathcal{L}}
\newcommand{\FLang}{\mathcal{F}}
\newcommand{\Mon}{\mathcal{M}}
\newcommand{\contrapow}{\mathcal{Q}}

\newcommand{\Prop}{\mathsf{Prop}}

\newcommand{\Sem}[1]{\lb #1\rb}

\newcommand{\Pow}{\mathcal{P}}

\newcommand{\Land}{\bigwedge}

\newcommand{\Nat}{\mathbb{N}}

\newcommand{\Int}{\mathbb{Z}}
\newcommand{\Reals}{\mathbb{R}}

\newcommand{\pb}{\mathsf{pb}}
\newcommand{\Up}{\mathsf{Up}}
\newcommand{\Nei}{\mathcal{N}}
\newcommand{\lb}{\llbracket}
\newcommand{\rb}{\rrbracket}

\newcommand{\by}[1]{(\text{#1})}

\renewcommand{\ell}{\mathsf{l}}
\newcommand{\Set}{\mathsf{Set}}

\newcommand{\hearts}{\heartsuit}

\newcommand{\set}{\mathsf{Set}}
\newcommand{\fun}{\mathsf{F}}
\newcommand{\bbX}{\mathbb{X}}

\theoremstyle{plain}
\newtheorem{thm}{Theorem}[section]
\newtheorem{lem}[thm]{Lemma}

\newtheorem{lemdefn}[thm]{Lemma and Definition}
\newtheorem{fact}[thm]{Fact}
\newtheorem{propn}[thm]{Proposition}

\theoremstyle{definition}
\newtheorem{defn}[thm]{Definition}
\newtheorem{expl}[thm]{Example}

\newtheorem{rem}[thm]{Remark}

\newcounter{blubber}
\newenvironment{sparenumerate}
{\begin{list}
  {\arabic{blubber}.}
  {\usecounter{blubber}
   \setlength{\leftmargin}{0pt}
    \setlength{\parsep}{0pt}
    \setlength{\itemindent}{4ex}
    \setlength{\itemsep}{2pt}
  }
}
{\end{list}}

\usepackage{microtype}%if unwanted, comment out or use option "draft"

\title{Uniform Interpolation in Coalgebraic Modal Logic\footnote{Work by the first and second author forms part of DFG project GenMod3 (SCHR 1118/5-3)}}
\titlerunning{Uniform Interpolation in Monotone Coalgebraic Modal Logic} %optional, in case that the title is too long; the running title should fit into the top page column

\author[1]{Fatemeh Seifan}
\author[1]{Lutz Schröder}
\author[2]{Dirk Pattinson}
\affil[1]{Friedrich-Alexander-Universität Erlangen-Nürnberg}
\affil[2]{Australian National University}
\authorrunning{F.\ Seifan, L.\ Schröder and D.\ Pattinson} %mandatory. First: Use abbreviated first/middle names. Second (only in severe cases): Use first author plus 'et. al.'

\Copyright{}

%Dummy classification -- please refer to \url{http://www.acm.org/about/class/ccs98-html}}% mandatory: Please choose ACM 1998 classifications from http://www.acm.org/about/class/ccs98-html . E.g., cite as "F.1.1 Models of Computation". 
% mandatory: Please provide 1-5 keywords
\begin{document}

\maketitle

\begin{abstract}
  % A logic has \emph{(Craig) interpolation} if each logical consequence
  % between two formulas is mediated by a formula -- the
  % \emph{interpolant} -- that mentions only the shared symbols. The
  % stronger property of \emph{uniform interpolation} requires
% additionally that the interpolant 
A logic has \emph{uniform interpolation} if its formulas can be
projected down to given subsignatures, preserving all logical
consequences that do not mention the removed symbols; the weaker
property of \emph{(Craig) interpolation} allows the projected
formula -- the \emph{interpolant} -- to be different for each
logical consequence of the original formula. These properties are of
importance, e.g., in the modularization of logical theories. We
study interpolation in the context of coalgebraic modal logics,
i.e.\ modal logics axiomatized in rank~$1$, restricting for clarity
to the case with finitely many modalities. Examples of such logics
include the modal logics $K$ and $KD$, neighbourhood logic and its
monotone variant, finite-monoid-weighted logics, and coalition
logic. We introduce a notion of one-step (uniform) interpolation,
which refers only to a restricted logic without nesting of
modalities, and show that a coalgebraic modal logic has uniform
interpolation if it has one-step interpolation. Moreover, we
identify preservation of finite surjective weak pullbacks as a
sufficient, and in the monotone case necessary, condition for
one-step interpolation. We thus prove or reprove uniform
interpolation for most of the examples listed above.
\end{abstract}

\section{Introduction}\label{sec:intro}

Given a logic with a notion of formula and signature (and featuring
implication for simplicity), the \emph{Craig interpolation} property
requires that every valid implication $\phi\to\psi$ has an
\emph{interpolant}, i.e.\ a formula $\rho$ mentioning only the
signature symbols that occur in both $\phi$ and $\psi$, such that both
$\phi\to\rho$ and $\rho\to\psi$ are valid. The stricter \emph{uniform
  interpolation} property additionally demands that $\rho$ can be made
to depend only on $\phi$ and on the signature of $\psi$ (or, yet
stricter, on the shared symbols of $\phi$ and $\psi$), rather than
also on $\psi$ itself. Both Craig interpolation and uniform
interpolation are useful in the structuring and modularization of
logical theories for purposes of specification and automated
deduction, e.g.\ in large
ontologies~\cite{WangEA09,LutzWolter11}. Craig interpolation was
originally proved for first-order logic~\cite{Craig57} and later
extended to many other systems, notably various modal logics including
the basic modal logic~$K$~\cite{Gabbay72}, as well as intuitionistic
logic~\cite{Gabbay77} and the $\mu$-calculus \cite{DagostinoH00}.
% Gabbay proves interpolation for modal predicate logics without
% equality This actually implies interpolation for the propositional
% case: The propositional case is the case with only nullary
% predicates; quantifiers can then just be removed because no formula
% mentions any variables. 
Uniform interpolation is easily seen to hold for classical
propositional logic but in fact fails for first-order predicate
logic~\cite{Henkin63}. Intuitionistic logic~\cite{Pitts92}, the basic
modal logic~$K$~\cite{Ghilardi95,Visser96}, and the modal
$\mu$-calculus~\cite{Kozen83} do have uniform interpolation, while it
fails for the modal logics $S4$~\cite{GhilardiZawadowski95} and
$K4$~\cite{Bilkova07}.

In this paper, we study interpolation and uniform interpolation in the
context of predicate-lifting style coalgebraic modal
logic~\cite{Pattinson04,Schroder08}, equivalently, of modal logics
that are axiomatized by \emph{rank-$1$}
axioms~\cite{Schroder07,SchroderPattinson10d}. Coalgebraic modal logic
is a generic framework for modal logics whose semantics goes beyond
the standard relational world, and e.g.\ includes probabilistic,
game-based, neighbourhood-based, or weighted behaviour. It is
parametrized over the choice of a \emph{type functor} (in our setting,
on the category of sets), whose coalgebras play the role of
models. The name of the game in coalgebraic logic is to reduce
properties of the full modal logic to properties of the
\emph{one-step} logic, which restricts to formulas with exactly one
layer of modalities and is interpreted over very simple structures
that essentially capture the collection of successors of a single
state in a model. Following this paradigm, we identify a notion of
one-step interpolation, and then establish that for a coalgebraic
modal logic~$\Lang$ with finitely many modalities, the following
properties imply each other in sequence
\begin{enumerate}
\item\label{item:pb} the modalities are \emph{separating}, i.e.\
  support a Hennessy-Milner-style expressivity
  theorem~\cite{Pattinson04,Schroder08} (implying that the type
  functor preserves finite sets), and the type functor preserves
  surjective finite weak pullbacks (which for finitary functors just
  means that the functor preserves surjective weak pullbacks);
\item $\Lang$ has one-step interpolation;
\item $\Lang$ has uniform interpolation.
\end{enumerate}
Here a pullback is called surjective if it consists of surjective
maps. If the modalities of $\Lang$ are separating and monotone, then
preservation of finite surjective weak pullbacks is in fact necessary
for one-step interpolation.

As applications of this result, we obtain that neighbourhood logic
(i.e.\ classical modal logic~\cite{Chellas80}), monotone modal
(neighbourhood) logic~\cite{Chellas80}, the relational modal logics
$K$ and $KD$, coalition logic~\cite{Pauly02}, and logics of
monoid-weighted transition systems for finite refinable commutative
monoids (in particular for finite Abelian groups, even though the
latter fail to admit monotone modalities) have uniform interpolation;
for neighbourhood logic, coalition logic, and monoid-weighted logics,
these results appear to be new. 

Proofs are often omitted or only sketched; full proofs are in the appendix. 

\subsubsection*{Related Work} Craig interpolation for monotone modal
logic was first proved by Hansen and Kupke~\cite{HansenKupke04} and
later improved to uniform interpolation by Santocanale and
Venema~\cite{SantocanaleVenema10}. Craig interpolation (but not
uniform interpolation) for coalition logic was proved by Schröder and
Pattinson using coalgebraic cutfree sequent
systems~\cite{PattinsonSchroder10}. Hansen, Kupke, and
Pacuit~\cite{HansenEA09} have proved Craig interpolation (but not
uniform interpolation) for neighbourhood logic, using semantic
methods.  Uniform interpolation for coalgebraic modal logic with a
generalized Moss modality based on a \emph{quasifunctorial lax
  lifting} has been shown, for functors preserving finite sets, by
Marti in his MSc thesis~\cite{MartiMSc} (and in fact this result has
been extended to coalgebraic modal fixpoint
logics~\cite{MartiEA15}). Logics based on diagonal-preserving lax
liftings (even without assuming quasi-functoriality) satisfy an
obvious variant of separation and thus support a generalized
Hennessy-Milner theorem, and moreover can be translated into the
language of monotone predicate liftings~\cite{MartiMSc}. We leave it
as an open problem to determine the relationship between
quasifunctoriality and preservation of surjective weak pullbacks in
presence of a separating set of monotone predicate liftings. We
emphasize that our criteria for interpolation apply also to logics
that fail to be separating or admit monotone modalities, hence cannot
be phrased in terms of quasifunctorial lax liftings, notable examples
of this type being coalition logic, neighbourhood logic, and logics of
finite-Abelian-group-weighted transition systems.

In~\cite{Pattinson13}, the (coalgebraic) logic of exact covers was
introduced; besides a generic Hennessy-Milner theorem and results on
completeness and small models, a generic uniform interpolation theorem
was claimed which implies that every rank-$1$ modal logic with
finitely many (not necessarily monotone) modalities has uniform
interpolation. We show by means of a counterexample that the latter
claim is incorrect; our results help delineate in which cases it can
be salvaged.

This paper is an extended version of a previous conference
publication~\cite{SeifanEA17}.

\section{Preliminaries}\label{sec:prelims}
We assume basic familiarity with category theory
(see~\cite{AdamekHerrlich90} for an introduction). Throughout, we work
over the category $\set$ of sets and functions as the base
category. Given a functor $\fun:\set\to\set$, an
\emph{$\fun$-coalgebra} is a pair $\bbX=(X, \xi)$ consisting of a
set $X$ (of states) and a function $\xi: X\to \fun X$. In the
spirit of coalgebraic logic, we use such coalgebras as generic models
of modal logics; e.g.\ Kripke frames can be seen as a coalgebras
$\xi:X\to\Pow X$ for the powerset functor $\Pow$, as they assign to
each state $x\in X$ a set $\xi(x)\in\Pow(X)$ of successor
states. We will later see non-relational examples.  % Given two
% $\fun$-coalgebras $\bbX=(X, \xi)$ and $\bbX '=(X', \xi ')$, a
% \emph{coalgebra morphism} $f: \bbX\to \bbX '$ is a function
% $f: X\to X'$ such that $\fun f\circ \xi =\xi '\circ f $.
We denote by $\contrapow$ the contravariant powerset functor, which
acts on sets by taking powersets and on maps by taking preimage maps
($\contrapow f(A)=f^{-1}[A]$).

\subsection{Set Functors and Weak Pullbacks}
 
The pullback of a cospan $(f,g)=X\xrightarrow{f} Z\xleftarrow{g}Y$ in
$\Set$ is described as
\begin{equation*}
  (\pb(f,g), \pi_{1}, \pi_{2})\ \text{where}\ \pb(f,g)=\{(x,y)\in X\times Y\mid f(x)=g(y)\}
\end{equation*}
and $\pi_{1}$, $\pi_{2}$ are the projections to the components.

An important property of set functors in the analysis of coalgebras is
weak pullback preservation~\cite{Rutten00}. A set functor $\fun$
\emph{preserves weak pullbacks}, or \emph{weakly preserves pullbacks}
if it maps pullbacks to \emph{weak pullbacks} (equivalently maps weak
pullbacks to weak pullbacks), where a weak pullback is defined
categorically like a pullback but requiring only existence, not
uniqueness, of mediating morphisms. In an element-wise formulation,
$X\xleftarrow{\pi_1}P\xrightarrow{\pi_2}Y$ is a weak pullback of
$X\xrightarrow{f} Z\xleftarrow{g}Y$ if whenever $f(x)=g(y)$ for
$x\in X$, $y\in Y$, then there exists $p\in P$ (not necessarily
unique) such that $\pi_1(p)=x$, $\pi_2(p)=y$. It well-known and easy
to see that the identity functor, all constant functors, and the
powerset functor preserve weak pullbacks and that the class of
weak-pullback preserving functors is closed under products,
coproducts, exponentiation with constants, functor composition, and
taking finitary parts~\cite{Rutten00,TuriThesis}. In particular, all
generalized Kripke polynomial functors (built from powerset, finite
powerset, constant functors, and identity by taking products,
coproducts, and exponentiation with constants) preserve weak
pullbacks. Some negative examples are as follows.
\begin{expl} \label{ex.counterex}\begin{sparenumerate}
  \item \label{ex:N}The \emph{Neighbourhood functor} or double
    contravariant powerset functor $\Nei=\contrapow\contrapow$ maps a
    set $X$ to $\Nei X = \contrapow\contrapow X$ and a function
    $f:X\to Y$ to
    $\Nei f(\alpha)=\{ A\subseteq Y \mid f^{-1}[A]\in\alpha\}$. This
    functor does not preserve weak pullbacks \cite{Rutten00}.
  \item \label{ex:M} A variant of the neighbourhood functor is the
    \emph{monotone neighbourhood functor} denoted by $\Mon$. Given an
    element $\alpha\in \contrapow\contrapow X$ we define
    $\mathsf{Up}(\alpha)$ to be the set
    \begin{equation*}
      \mathsf{Up}(\alpha):=\{Y\subseteq X\mid Y\supseteq Z\
      \text{for some}\ Z\in \alpha\},
    \end{equation*}
    and we say that $\alpha$ is \emph{upwards closed} if
    $\alpha=\mathsf{Up}(\alpha)$.  The functor $\Mon$ is then given on
    sets $X$ by
    $ \Mon X = \{\alpha\in\contrapow\contrapow(X)\mid \alpha\text{
      upwards closed}\}$
    and on maps $f:X\to Y$ by
    $\Mon f(\alpha) = \{A\subseteq Y\mid f^{-1}[A]\in\alpha\} =
    \contrapow\contrapow(f)(\alpha)$.
    Like $\Nei$, $\Mon$ does not preserve weak
    pullbacks~\cite{HansenKupke04}.

  \item Another functor that does not preserve weak pullbacks is
    $\fun^{3}_{2}$, defined as a subfunctor of the cubing functor
    $X\mapsto X^3$ by
    $\fun^{3}_{2} X= \{(x_1, x_2, x_3)\in X^{3}\mid\; |\{x_1, x_2,
    x_3\}|\leq 2\}$~\cite{GummSchroder00}.
  \end{sparenumerate}
\end{expl}

\subsection{Coalgebraic Modal Logic}\label{sec:cml}
We briefly recall the syntax and semantics of coalgebraic modal logic.

We fix a countable set $V$ of \emph{propositional variables}. The
syntax of a coalgebraic modal logic $\Lang(\Lambda)$ is then
determined by the choice a \emph{modal signature} $\Lambda$ consisting
of modal operators with assigned arities: the set $\FLang(\Lambda)$ of
\emph{(modal) $\Lambda$-formulas} is defined by the grammar
\[\Lang(\Lambda)\ni\phi,\psi::=v\mid\bot\mid\neg
\phi\mid\phi\wedge\psi\mid \hearts(\phi_{1}, \cdots, \phi_{n}),
\end{equation*}
where $v\in V$ and $\hearts\in \Lambda$ is an $n$-ary modality (we
deviate slightly from usual practice in coalgebraic modal logic by
including propositional variables in the syntax rather then regarding
them as nullary modalities; this is in order to facilitate the
definition of interpolation). Other Boolean operators
($\top, \lor, \to, \leftrightarrow $) are defined in the standard
way. We write
$\rank(\phi)$ for the \emph{rank} of $\phi$, i.e. the maximal nesting
depth of modal operators in $\phi$.

As indicated above, the type of systems underlying the semantics of
$\Lang(\Lambda)$ is then determined by the choice of a set functor
$\fun$ whose coalgebras play the role of frames. The interpretation of
the modal operators is then defined in terms of predicate liftings for
$\fun$:
\begin{defn}[Predicate liftings] \label{def:PredicateLifting} An
  \textit{$n$-ary predicate lifting} for $\fun$ is a natural
  transformation
  $\lambda : \contrapow(-)^n \rightarrow \contrapow \circ \fun$ ,
  where $\contrapow(-)^n$ denotes the $n$-fold product of $\contrapow$
  with itself. We say that $\lambda$ is \textit{monotone} if
  $ \lambda_X(Y_1,...,Y_n) \subseteq \lambda_X(Z_1,...,Z_n), $
  whenever $Y_i \subseteq Z_i\subseteq X$ for each $i$. Equivalently, we
  can describe $\lambda$ by its \emph{transposite}
  $\lambda^{\flat}: \fun\to \contrapow\contrapow^{n}$ given by
  $\lambda^{\flat}_{X}(t)=\{(Y_1,...,Y_n)\in\contrapow\contrapow^{n}X\mid
  t\in \lambda_{X}(Y_1,...,Y_n)\}$.
\end{defn}
\noindent By the Yoneda lemma, we have the following equivalent
description of predicate liftings~\cite{Schroder08}.

\begin{fact} The $n$-ary predicate liftings for $\fun$ are in
  one-to-one correspondence with subsets of $\fun(2^n)$, where
  $2=\{\top, \bot\}$; for $n=1$, such a subset $U$ determines a
  predicate lifting $\lambda$ by
  $\lambda_X(A)=\{t\in\fun X\mid T\chi_A(t)\in U\}$ where
  $\chi_A:X\to 2$ is the characteristic map of $A\subseteq X$.
\end{fact}  

\noindent We then complete the semantic parametrization of
$\Lang(\Lambda)$ by assigning to each $n$-ary modal operator
$\hearts\in\Lambda$ an $n$-ary predicate lifting $\Sem{\hearts}$ for
$\fun$. For readability, we mostly restrict the technical exposition
to unary modalities from now on; the extension to finitary modalities
is just a matter of adding indices.
\begin{defn}\label{def:semantics} An \emph{$\fun$-model} $(X,\xi, \tau)$ 
  consists of an $\fun$-coalgebra $\bbX=(X,\xi)$ and a valuation
  $\tau:V\to \mathcal{P}(X)$ of the propositional variables.  We then
  inductively define a satisfaction relation $\vDash$ between states
  of the model $(X,\xi, \tau)$ and formulas of $\Lang(\Lambda)$ by
  $x\Vdash v$ iff $x\in \tau(v)$, standard clauses for Boolean
  connectives, and
  \begin{equation*}
    x\vDash \hearts(\phi_1,...,\phi_n)\ \text{ iff}\  \xi(x)\in 
    \hearts_X(\lb \phi_1 \rb,...,\lb \phi_n\rb),
  \end{equation*}
  where $\lb\phi_i \rb = \{t \mid t\vDash \phi_i\}$. As usual, we say
  that a formula $\phi$ is \emph{satisfiable} if there exists a state
  $x$ in some model such that $x\models\phi$, and \emph{valid} if
  $x\models\phi$ for every state $x$ in every $\fun$-model.
\end{defn}

\begin{expl}\label{ex:predicateliftings}
   \begin{enumerate}\item
     The modal logic $K$ is captured coalgebraically by taking the
     powerset functor $\Pow$ as the type functor,
     $\Lambda=\{\Diamond\}$, and
     \begin{equation*}\Sem{\Diamond}_{X}(Y)=\{A\in \Pow X\mid A\cap
       Y\neq\emptyset\}.
     \end{equation*}
     % \begin{equation*}\alpha \in \Diamond_X (Z) \Leftrightarrow \alpha \cap Z \neq \emptyset.\end{equation*}

   \item Neighbourhood logic (or \emph{classical modal
       logic}~\cite{Chellas80}) has $\Lambda=\{\Box\}$, interpreted
     over the neighbourhood functor $\Nei$
     (Example~\ref{ex.counterex}.\ref{ex:N}) by
     \begin{equation*}\Sem{\Box}_{X}(Y):=\{\alpha\in \Nei X\mid
       Y\in\alpha\}.
     \end{equation*}
     \emph{Monotone modal (neighbourhood) logic} is captured in the
     same way, replacing $\Nei$ with the monotone neighbourhood
     functor $\Mon$ (Example~\ref{ex.counterex}.\ref{ex:M}).
\end{enumerate}
\end{expl}\medskip

\noindent \emph{We fix the data $\fun$, $\Lambda$, $\Sem{\hearts}$ of
  the logic $\Lang(\Lambda)$ from now on.}

\myparagraph{The One-Step Logic} Given any set $Z$, we denote by
$\Prop(Z)$ the set of propositional formulas over $Z$:
\begin{equation*}
  \Prop(Z)\owns\phi::= \bot\mid z\mid\neg\phi\mid\phi\land\phi\qquad(z\in Z),
\end{equation*}
and write $\Lambda(Z)$ for the set of formulas
$\hearts(z_{1},\cdots ,z_{n})$ where $\hearts \in \Lambda$ has arity
$n$ and $z_{1}, \cdots , z_{n}\in Z$. We then define a \emph{one-step
  formula over $Z$} to be an element of $\Prop(\Lambda(\Prop(Z)))$.
Here, $Z$ will often be a subset of $V$; also, $Z$ will sometimes be a
subset of some powerset $\Pow X$, in which case we will understand
every element of $\Pow X$ to be interpreted as itself. In general, we
interpret both propositional formulas and one-step formulas over $Z$
w.r.t.\ $\Pow(X)$-valuations $\tau: Z\to \Pow(X)$ for some set $X$: We
extend $\tau$ to propositional formulas using the Boolean algebra
structure of $\Pow X$, obtaining for $\phi\in\Prop(Z)$ a subset
\begin{equation*}
  \phi\tau\in\Pow X.
\end{equation*}
We write $X\models\phi\tau$ if $\phi\tau=X$. We then define the
extension
\begin{equation*}
\psi\tau\in\Pow( \fun X)
\end{equation*}
of a one-step formula $\psi\in\Prop(\Lambda(\Prop(Z)))$ recursively by
the evident clauses for Boolean connectives, and
\begin{equation*}
  (\hearts\phi)\tau = \Sem{\hearts}_X(\phi\tau).
\end{equation*}
When $Z\subseteq\Pow(X)$ and $\tau$ is just subset inclusion, the we
omit $\tau$ from the notation, so $\psi\in\Prop(\Lambda(\Pow X))$
denotes both a one-step formula and its interpretation in
$\Pow(\fun X)$.
\begin{defn}
  A one-step formula $\psi\in\Prop(\Lambda(\Prop(Z)))$ is
  \emph{(one-step) satisfiable over $\tau:Z\to\Pow(X)$} if
  $\psi\tau\neq\emptyset$, and \emph{(one-step) satisfiable} if $\psi$
  is one-step satisfiable over $\tau$ for some~$\tau$. Dually, $\psi$
  is \emph{(one-step) valid (over $\tau$)} if $\neg\psi$ is (one-step)
  unsatisfiable (over~$\tau$). We write $\fun X,\tau\models\psi$ if $\psi$
  is one-step valid over $\tau$, and $\models\psi$ if $\psi$ is
  one-step valid.
\end{defn}
We will need the following pieces of terminology and notation:
\begin{defn}\label{def:invariance}
  For a map $f:X\to Y$, we write $\sigma_f$ for the
  \emph{substitution} mapping $A\in\Pow(X)$ to $f[A]$ (e.g.\ in
  one-step formulas of type $\Prop(\Lambda(\Pow(X)))$), and
  $\sigma_{f^-}$ for the substitution mapping $B\in\Pow(Y)$ to
  $f^{-1}[B]$.  A set $A\in\Pow(X)$ is \emph{$f$-invariant} if
  $f^{-1}[f[A]]=A$.
\end{defn}
Clearly, all sets of the form $f^{-1}[B]$ are $f$-invariant, i.e.\ the
$f$-invariant sets are precisely those of the form $f^{-1}[B]$. The
$f$-invariant sets form a Boolean subalgebra of $\Pow(X)$.

\begin{defn}
  We denote by $\Ats(\BA)$ the set of \emph{atoms} of a finite Boolean
  algebra $\BA$, i.e.\ its minimal non-bottom elements, and $\can_\BA$ for
  the canonical isomorphism $\BA\to\Pow(\Ats(\BA))$. Given a
  subalgebra $\BA_0$ of $\BA$, we have a \emph{canonical projection}
  $\Ats(\BA)\to\Ats(\BA_0)$. 
  % Maps A to the intersection B of all BA_0-elements above A.
  % This is an atom: it is non-bottom, and if there is some C below it
  % then it is not above A, so -C above A
  % and then B is in -C, contradiction.
\end{defn}

%  \begin{defn}[Canonical valuation] We denote by
%  $\ev_Z:Z\to\Pow(2^Z)$ the \emph{canonical $\Pow(2^Z)$-valuation}
%  given by
%  \begin{equation*}
%  \ev_Z(a)=\{\kappa\in 2^Z\mid \kappa(a)=\top\}.
% \end{equation*}
% \end{defn}
\noindent The following lemmas are straightforward consequences of
naturality of predicate liftings:
\begin{lem} \label{lem:can-mod} Given a finite Boolean subalgebra
  $\BA$ of $\Pow X$ for a set $X$, $\phi\in\Prop(\Lambda(\BA))$ is
  satisfiable (valid) iff $\phi\can_\BA$ is satisfiable (valid).
\end{lem}

% \begin{lem}~\cite[Proposition~23]{Schroder07} \label{lem:can-mod} A
%   one-step formula $\phi\in\Prop(\Lambda(\Prop(Z)))$ is satisfiable
%   (valid) iff it is satisfiable (valid) over the canonical valuation
%   $\ev_Z$.
% \end{lem}
\begin{lem}\label{lem:restriction}
  Let $\BA_0\subseteq \BA_1$ be finite Boolean subalgebras of $\Pow X$
  for a set $X$, let $f:\Ats(\BA_1)\to\Ats(\BA_0)$ be the canonical
  projection, let $\phi\in\Prop(\Lambda(\BA_0))$, and let
  $t\in \fun(\Ats(\BA_1))$.  Then $t\in\phi\can_{\BA_1}$ iff
  $\fun f(t)\in\phi\can_{\BA_0}$.
\end{lem}
% \begin{lem}\label{lem:restriction}
%   Let $V_0\subseteq V_1$ be sets of propositional variables, and let
%   $f:2^{V_1}\to 2^{V_0}$ act by restriction. Let
%   $\phi\in\Prop(\Lambda(\Prop(V_0)))$, and let $t\in \fun(2^{V_1})$. Then
%   $t\in\phi\ev_{V_1}$ iff
%   $\fun f(t)\in\phi\ev_{V_0}$. 
% \end{lem}

\myparagraph{Separation and Maximally Satisfiable Sets} The key
condition ensuring that $\Lang(\Lambda)$ satisfies the Hennessy-Milner
property, i.e.\ distinguishes non-bisimilar states, is
\emph{separation}~\cite{Pattinson04,Schroder08}:
\begin{defn}[Separation]
  We say that $\Lambda$ is \emph{separating} if for each set $X$, the
  family of maps
  $(\Sem{\hearts}^{\flat}_{X}: \fun X\to\contrapow\contrapow X=\Nei
  X)_ {\hearts\in\Lambda}$
  is jointly injective. % This means that every $t\in \fun X$ is
  % uniquely determined by the set
  % $\{(\lambda, A)\mid\lambda\in \Lambda, A\in \contrapow(X),t\in
  % \lambda_{X}(A)\}$.
\end{defn}
We proceed to define the \emph{MSS-functor} (for \emph{maximally
  one-step satisfiable sets}) from $\fun$ and $\Lambda$. (A related
functor using \emph{maximally one-step consistent} sets has been used
to show that every rank-$1$ modal logic has a coalgebraic
semantics~\cite{SchroderPattinson10d}.)
   
\begin{defn} A set $\Phi\subseteq\Prop(\Lambda(\Pow(X)))$ is
  \emph{one-step satisfiable} if the intersection of the
  interpretations of the formulas in $\Phi$ is non-empty, and
  \emph{maximally one-step satisfiable} if $\Phi$ is maximal among
  such sets. The \emph{MSS-functor} $M^\Lambda_\fun$ is given by
  $M^\Lambda_{\fun} X$ being the set of maximally one-step satisfiable
  subsets of $\Prop(\Lambda(\Pow X))$, and
  $M^\Lambda_{\fun}f(\Phi)=\{\phi\in\Prop(\Lambda(\Pow(Y)))\mid\phi\sigma_{f^-}\in\Phi\}$
  for $f:X\to Y$.
\end{defn} 
The following lemma allows us to identify $\fun$ with its MSS-functor
whenever $\Lambda$ is separating.

\begin{lem}\label{lem:mss} If $\Lambda$ is separating, then $\fun$ and
  $M^\Lambda_{\fun}$ are isomorphic.
\end{lem}

\section{Surjective Weak Pullbacks}\label{sec:pullbacks}
We proceed to introduce the key semantic interpolation criterion,
preservation of \emph{surjective} weak pullbacks. We record
explicitly:
 \begin{defn}
  A pullback of a cospan $(f,g)$ of maps (in $\Set$) is
  \emph{surjective} if both $f$ and $g$ are surjective, and
  \emph{finite} if all involved sets are finite. A functor
  \emph{preserves (finite) surjective weak pullbacks} if it maps
  (finite) surjective pullbacks to weak pullbacks.
\end{defn}
\noindent Recall that under AC, every set functor preserves surjective
maps. Also, surjective maps are stable under pullbacks, so all
morphisms in a surjective pullback are surjective.  non-empty binary
Cartesian products $X\times Y$ are surjective pullbacks of
$X\to 1\leftarrow Y$.  % ; for
% clarity, we call such pullbacks \emph{single-fibre pullbacks}, and say
% that $\fun$ \emph{preserves (finite) single-fibre weak pullbacks} if
% for every (finite) $X\times Y\neq\emptyset$, $\fun$ preserves the
% surjective weak pullback $X\times Y$ of $X\to 1\leftarrow Y$ 
Moreover, the kernel pair of a map $f:X\to Y$ is a surjective pullback
of the codomain restriction $X\to f[X]$.

For finitary functors, the finiteness restriction in the preservation
condition is immaterial:
\begin{lem}\label{lem:finitary-pb}
  If $\fun$ is finitary, then $\fun$ preserves (surjective) weak pullbacks
  iff $\fun$ preserves finite (surjective) weak pullbacks. 
\end{lem}
Of course, every functor that preserves weak pullbacks also preserves
surjective weak pullbacks, e.g.\ the (finite or unrestricted) powerset
functor, and more generally all Kripke polynomial functors. Two
negative examples are as follows.

\begin{expl} \label{expl:no-surj-weak-pb}\begin{enumerate}
  \item\label{item:nbhd} The neighbourhood functor $\Nei$ fails to
    preserve finite surjective weak pullbacks. To see this, consider
    the pullback of the following functions as in \cite{Rutten00}. Let
    $X=\{a_{1}, a_2, a_3\}$, $Y=\{b_1, b_2, b_3\}$ and
    $Z=\{c_1, c_2\}$ and define surjective maps $f: X\to Z$ and
    $g:Y\to Z$ as follows: $f(a_1)=f(a_2)=c_1$, $f(a_3)=c_2$,
    $g(b_1)=c_1$ and
    $g(b_2)=g(b_3)=c_2$.% For $u=\{\{a_1\}\}$ and $v=\{\{b_3\}\}$
  %it is impossible to find a $w\in\Nei\pb(f,g)$ such $\Nei\pi_{1}(w)=u$ and $\Nei\pi_{2}(w)=v$. 
 % \item The functor $\fun^{3}_{2}$ fails to preserve finite surjective
  %  weak pullbacks. For a counterexample, take $f:X\to Z$ as in the
  %  case of the neighbourhood functor. It is not difficult to
  %  calculate that $|\pb(\fun^{3}_2 f, \fun^{3}_2 f)|=125$ and
  %  $|\fun^{3}_2 \pb(f, f)|=65$, hence there is no surjective map from
  %  $\fun^{3}_2(f, g)$ to $\pb(\fun^{3}_2 f, \fun^{3}_2 g)$; by
 %   Fact~\ref{fact:WPb} this implies that $\fun^{3}_2$ does not weakly
  %  preserve the pullback of $(f,f)$.
\item The functor $\fun^{3}_{2}$ fails to preserve finite surjective
  weak pullbacks. For a counterexample consider a surjective cospan
  $(f, g)$ with $f=g$ being the constant map $\{a, b\}\to \{b\}$.  For
  $u=(b, b, a)$ and $v=(a, b, b)$, it is impossible to find a
  $w\in\fun^{3}_{2}\pb (f,g)$ such that $\fun^{3}_{2}\pi_{1}(w)=u$ and
  $\fun^{3}_{2}\pi_{2}(w)=v$.  % As shown by Gumm and Schr{\"o}der \cite{GummSchroder00}, $\fun^{3}_{2}$ weakly preserves pullback of $(f, g)$ if at least one of $f$, $g$ is injective.
    
\end{enumerate}
\end{expl}

\noindent We proceed to see examples that fail to preserve weak
pullbacks but do preserve surjective weak pullbacks.

\paragraph*{The Monotone Neighbourhood Functor}

The monotone neighbourhood functor $\Mon$ does not preserve all weak
pullbacks (Example~\ref{ex.counterex}.\ref{ex:M}). However:

\begin{propn}\label{prop:mon} 
  The monotone neighbourhood functor $\Mon$ preserves surjective weak
  pullbacks.
\end{propn}
\noindent The proof is facilitated by the following fact:
\begin{lemdefn}[Compatibility]\label{lem:compatible}
Let
  \begin{equation}\label{eq:pb}
    \begin{tikzcd}
      P \arrow{d}[left]{\pi_2}\arrow{r}{\pi_1} & X\arrow{d}{f}\\
      Y \arrow{r}[below]{g} & Z
    \end{tikzcd}
  \end{equation}
  be a surjective pullback, and let $\alpha_1\in\Mon X$,
  $\alpha_2\in\Mon Y$. Then $\Mon f(\alpha_1)=\Mon g(\alpha_n)$ iff
  $\alpha_1$ and $\alpha_2$ are \emph{compatible}, i.e.\ for every
  $U\in\alpha_1$ we have $\pi_2[\pi_1^{-1}[U]]\in\alpha_2$ and
  symmetrically.
\end{lemdefn}

\begin{proof}[Proof (Proposition~\ref{prop:mon}, Sketch)]
  Given a surjective pullback~\eqref{eq:pb} and compatible
  $\alpha_1\in\Mon X$, $\alpha_2\in\Mon Y$, it is straightforward to
  show that
  \begin{equation*}
    \beta=\Up(\{\pi_{1}^{-1}[U]\mid U\in \alpha_1\}\cup\{\pi_{2}^{-1}[V]\mid V\in \alpha_{2}\})\in\Mon P
  \end{equation*}
  satisfies $\Mon\pi_{1}(\beta)=\alpha_{1}$ and
  $\Mon\pi_{2} (\beta)=\alpha_{2}$.
\end{proof}

\paragraph*{Monoid-weighted Functors}

Given a commutative monoid $M$ (which we write additively), the
\emph{monoid-weighted functor} $\Wfun{M}$ is defined by taking
$\Wfun{M}X$ to be the set of finitely supported functions $X\to M$
(i.e.\ functions that vanish almost everywhere), and
$\Wfun{M} f(\mu)=\lambda y.\,\sum_{f(x)=y}\mu(x)$ for $f:X\to Y$ and
$\mu\in\Wfun{M}X$. Examples of monoid-weighted functors include the
free Abelian groups functor ($M=\Int$), the free vector space functor
($M=\Reals$), the finite multiset functor ($M=\Nat$), and the finite
powerset functor ($M=2=\{\bot,\top\}$ with $+$ being disjunction).
\begin{defn}[Refinability]
  \cite{GummSchroder01} A commutative monoid $M$ is \emph{refinable}
  if whenever $\sum_{i=1}^na_i=\sum_{j=1}^kb_j$ for
  $a_1,\dots,a_n,b_1,\dots,b_k\in M$, $n,k\ge 1$, then there exists an
  $n\times k$-matrix over $M$ with row sums $a_i$ and column sums
  $b_j$.
\end{defn}
As shown by Gumm and Schröder~\cite{GummSchroder01}, $\Wfun{M}$
preserves weak kernel pairs iff $M$ is refinable. In fact,
refinability already ensures preservation of all weak surjective
pullbacks:
\begin{lem}\label{lem:refinable}
  The functor $\Wfun{M}$ preserves weak surjective pullbacks iff $M$
  is refinable.
\end{lem}
Given that a) weak pullback preserving finitary functors are known to
admit separating sets of monotone predicate
liftings~\cite{KurzLeal12}, and b) the monotone neighbourhood functor
itself preserves surjective weak pullbacks but not all weak pullbacks,
it is tempting to conjecture that preservation of surjective weak
pullbacks is already sufficient for existence of a separating set of
monotone predicate liftings. This is not true, however:
\begin{defn}
  A commutative monoid $M$ is \emph{positive} if for $a,b\in M$,
  $a+b=0$ implies $a=b=0$.
\end{defn}
\begin{propn}\label{prop:positive}
  Let $M$ be refinable. Then $\Wfun{M}$ has a separating set of
  monotone predicate liftings iff $M$ is positive. 
\end{propn}
That is, every commutative monoid that is refinable but not positive
gives rise to a monoid-weighted functor that preserves surjective weak
pullbacks but does not admit a separating set of monotone predicate
liftings. One class of such commutative monoids are the non-trivial
Abelian groups: they clearly fail to be positive, and are easily seen
to be refinable~\cite{GummSchroder01}.

\section{One-Step Interpolation}

We proceed to develop our notion of one-step interpolation, and its
relationship to preservation of surjective weak pullbacks. From here
on, we assume throughout that \emph{the modal signature~$\Lambda$ is
  finite}.

%We say that $\Lang(\Lambda)$ is \emph{one-step compact} if the
%one-step logic is compact, and \emph{finitely one-step compact} if the
%one-step logic becomes compact when restricted to any finite set of
%variables, equivalently if every $\Prop(\Lambda(\Pow(X)))$ for finite
%$X$ is compact.

%\section{One-Step Interpolation}

\begin{defn}\label{def:os-interpol}
  Two Boolean subalgebras $\BA_1$, $\BA_2$ of $\Pow(X)$ for a set $X$
  are \emph{interpolable} if whenever $A\subseteq B$ for $A\in\BA_1$
  and $B\in\BA_2$, then there exists $C\in\BA_1\cap\BA_2$ such that
  $A\subseteq C$ and $C\subseteq B$. We say that $\Lang(\Lambda)$ has
  \emph{one-step interpolation} if given interpolable $\BA_1$, $\BA_2$
  and $\phi\in\Prop(\Lambda(\BA_1))$, $\psi\in\Prop(\Lambda(\BA_2))$
  such that $\fun X\models\phi\to\psi$, there is always an
  \emph{interpolant} $\rho\in\Prop(\Lambda(\BA_1\cap\BA_2))$ such that
  $\fun X\models\phi\to\rho$ and $\fun X\models\rho\to\psi$. Moreover,
  $\Lang(\Lambda)$ has \emph{uniform one-step interpolation} if the
  interpolant can be made to depend only on $\phi$ and
  $\BA_1\cap\BA_2=:\BA_0$; it is then called a \emph{uniform
    $\BA_0$-interpolant} of $\phi$.
\end{defn}
\noindent It is in fact not hard to see that under our running
assumptions, these two notions coincide, so refer to them just as
\emph{one-step interpolation}:
\begin{lem}\label{lem:os-unif-interpol}
  The logic $\Lang(\Lambda)$ has one-step uniform interpolation iff
  $\Lang(\Lambda)$ has one-step interpolation.
\end{lem}
\begin{proof}
  `Only if' is trivial. For `if', the assumption implies that, given
  data as in Definition~\ref{def:os-interpol},
  \begin{equation*}
    i(\phi)=\Land\{\rho\in\Prop(\Lambda(\Prop(\BA_0)))\mid \fun X\models \phi\to\rho\},
  \end{equation*}
  effectively a finite formula because $\Lambda$ is finite, is a
  uniform $\BA_0$-interpolant of $\phi$.
\end{proof}
\begin{rem}\label{rem:interpol-simplify}
  In any logic supporting the requisite propositional connectives, the
  set of formulas $\phi$ having a uniform interpolant is easily seen
  to be closed under disjunction, and similarly the set of pairs of
  formulas $\phi,\psi$ such that an interpolant between $\phi$ and
  $\psi$ exists is closed under disjunction in $\phi$ and under
  conjunction in $\psi$. When establishing (one-step) uniform
  interpolation for $\phi$, or (one-step) interpolation between $\phi$
  and $\psi$, we can therefore assume that $\phi$ is a conjunctive
  clause over modalized formulas and that $\psi$ is a disjunctive
  clause over modalized formulas.
\end{rem}

\noindent Our first positive example is neighbourhood logic:
\begin{expl}\label{expl:nbhd-interpol}
  Neighbourhood logic has one-step interpolation (and hence, by
  Lemma~\ref{lem:os-unif-interpol}, uniform one-step
  interpolation). To see this, let $\BA_1,\BA_2$ be interpolable
  Boolean subalgebras of $\Pow(X)$, let $\phi$ be a conjunctive clause
  over $\Lambda(\BA_1)$, and let $\psi$ be a disjunctive clause over
  $\Lambda(\BA_2)$ such that $\fun X\models\phi\to\psi$ (this case
  suffices by Remark~\ref{rem:interpol-simplify}). We can assume
  w.l.o.g.\ that $\fun X\not\models\neg\phi$ and
  $\fun X\not\models\psi$. Then $\fun X\models \phi\to\psi$ implies
  that $\phi$ contains a conjunct $\epsilon\Box A$ and $\psi$ a
  disjunct $\epsilon\Box B$, with $\epsilon$ representing either
  nothing or negation, such that $A = B$. Then $\epsilon\Box A$
  interpolates between $\phi$ and $\psi$.
\end{expl}

\noindent Preservation of surjective weak pullbacks is sufficient for
uniform one-step interpolation:
\begin{lem}\label{lem:pres-implies-interpol}
  Let $\Lambda$ be separating, and let $\fun$ preserve finite
  surjective weak pullbacks. Then $\Lang(\Lambda)$ has one-step
  uniform interpolation.
\end{lem}
\begin{proof}
  Let $\BA_0\subseteq \BA_1$ be finite Boolean subalgebras of
  $\Pow X$, and let $\phi\in\Prop(\Lambda(\BA_1))$. We show that
  \begin{equation*}
    i(\phi)=\Land\{\rho\in\Prop(\Lambda(\BA_0))\mid 
    \fun X\models\phi\to\rho\}
  \end{equation*}
  (effectively a finite formula) is a uniform $\BA_0$-interpolant for
  $\phi$. Dually, we show for $\psi\in\Prop(\Lambda(\BA_2))$ with
  $\BA_1$, $\BA_2$ interpolable, $\BA_1\cap \BA_2\subseteq \BA_0$, and
  $i(\phi)\land\psi$ satisfiable that also $\phi\land\psi$ is
  satisfiable.
  % (For we clearly have $\models\phi\to i(\phi)$, and if
  % $\phi\to\psi$ for $\psi\in\Prop(\Lambda(\Prop(V_2)))$, then
  % $\phi\land\neg\psi$ is unsatisfiable, so by the above, so is
  % $i(\phi)\land\neg\psi$, and therefore $\models i(\phi)\to\psi$.)
  By Lemma~\ref{lem:can-mod}, we have $s\in \fun(\Ats(\BA_2))$ such that
  $s\models(i(\phi)\land\psi)\can_{\BA_2}$. Let $\theta$ be the
  $\Prop(\Lambda(\BA_1\cap\BA_2))$-theory
  \begin{math}
    \theta=\Land\{\rho\in\Prop(\Lambda(\BA_1\cap\BA_2))\mid s\models
    \rho\can_{\BA_2}\}
  \end{math}
  of $s$. Then $\phi\land\theta$ is satisfiable: otherwise,
  $\fun X\models\phi\to\neg\theta$, so
  $\fun X\models i(\phi)\to\neg\theta$ by definition of $i(\phi)$,
  which by Lemma~\ref{lem:can-mod} contradicts
  $s\models (i(\phi)\land\theta)\can_{\BA_2}$.

  Again by Lemma~\ref{lem:can-mod}, we thus have
  $t\in \fun(\Ats(\BA_1))$ such that
  $t\models(\phi\land\theta)\can_{\BA_1}$. Let $\BA$ be the Boolean
  subalgebra of $\Pow X$ generated by $\BA_1\cup\BA_2$. Then the diagram
  \begin{equation*}
    \begin{tikzcd}
      \Ats(\BA) \arrow{r}[above]{\pi_1}\arrow{d}[left]{\pi_2}
      & \Ats(\BA_1) \arrow{d}[right]{f}\\
      \Ats(\BA_2)\arrow{r}[below]{g}  & \Ats(\BA_1\cap\BA_2),
      % & 2^{V_1\cup V_2} \arrow{dl}[above]{\pi_1}\arrow{dr}{\pi_2}\\
      % 2^{V_1} \arrow{dr}[below]{f} & & 2^{V_2}\arrow{dl}{g}\\
      % & 2^{V_1\cap V_2},
    \end{tikzcd}
  \end{equation*}
  where all maps are canonical projections, is a finite surjective
  pullback because $\BA_1$, $\BA_2$ are interpolable, hence weakly
  preserved by $\fun$. We claim that $\fun f(t)=\fun g(s)$: indeed,
  both sides satisfy $\theta\can_{\BA_1\cap \BA_2}$ by
  Lemma~\ref{lem:restriction}, and since $\theta$ is a complete
  $\Prop(\Lambda(\BA_1\cap \BA_2))$-theory, equality follows by
  separation. It follows that we have $u\in \fun (\Ats(\BA))$ such
  that $\fun \pi_1(u)=t$ and $\fun \pi_2(u)=s$. Again by
  Lemma~\ref{lem:restriction}, $u\models\phi\can_{\BA}$ and
  $u\models\psi\can_{\BA}$, so by Lemma~\ref{lem:can-mod},
  $\phi\land\psi$ is satisfiable.
\end{proof}

\noindent The example of neighbourhood logic
(Examples~\ref{expl:nbhd-interpol}
and~\ref{expl:no-surj-weak-pb}.\ref{item:nbhd}) shows that the
converse of Lemma~\ref{lem:pres-implies-interpol} does not hold in
general. It does however hold in the monotone case:
\begin{lem}\label{lem:interpol-implies-preservation}
  Let $\Lang(\Lambda)$ be monotone and separating and have one-step
  interpolation. Then~$\fun$ preserves finite surjective weak
  pullbacks.
\end{lem}
\noindent The proof relies on invariant sets, and uses the following lemma: 
\begin{lem}\label{lem:invariance}
  Let $f:X\to Y$ be surjective, let $\BA$ denote the subalgebra of
  $\Pow(X)$ consisting of the $f$-invariant sets, and let
  $\phi\in\Prop(\Lambda(\BA))$. Then $TX\models\phi$ iff
  $TY\models\phi\sigma_f$.
\end{lem}

\begin{proof}[Proof (Lemma~\ref{lem:interpol-implies-preservation}, Sketch)]
  Let $X\xleftarrow{\pi_1} P\xrightarrow{\pi_2} Y$ be a finite
  surjective pullback of $X\xrightarrow{f}Z\xleftarrow{g}Y$ as
  in Diagram~\eqref{eq:pb}.
  % \begin{center}
  %   \begin{tikzcd}
  %     & R \arrow{dl}[above]{\pi_1}\arrow{dr}{\pi_2}\\
  %     X \arrow{dr}[below]{f} & & Y\arrow{dl}{g}\\
  %     & Z
  %   \end{tikzcd}
  % \end{center}
  % be a finite surjective
  % pullback. % ; we understand $R$ as a difunctional
  % % relation between $X$ and $Y$, in particular we assume
  % % $R\subseteq X\times Y$
  As indicated in Section~\ref{sec:prelims}, we can identify $\fun$ with
  its MSS functor, i.e.\ we assume that $\fun X$ consists of maximally
  satisfiable subsets
  $\Phi\subseteq\Prop(\Lambda(\Pow(X)))$. % Of course, $\Phi$ is
  % completely determined by its restriction to $\Lambda(\Pow(X))$. We
  % assume moreover that $\Lambda$ is closed under duals, and work with
  % negation normal forms when convenient.
  In this reading, we are given $\Phi_1\in \fun X$ and
  $\Phi_2\in \fun Y$ such that $\fun f(\Phi_1)=\fun g(\Phi_2)$, which
  by a straightforward generalization of Lemma~\ref{lem:compatible}
  means that $\Phi_1$ and $\Phi_2$ are \emph{compatible}, i.e.\
  \begin{equation*}
    \phi\in\Phi_2\text{ implies }\phi\sigma_{\pi_2^-}\sigma_{\pi_1}\in\Phi_1
  \end{equation*}
  and symmetrically. We have to show that there exists
  $\Phi\in \fun R$ such that $F\pi_1(\Phi)=\Phi_1$ and
  $F\pi_2(\Phi)=\Phi_2$, i.e.\
  \begin{equation}\label{eq:project}
    \phi\in\Phi_1\iff \phi\sigma_{\pi_1^-}\in\Phi
  \end{equation}
  and correspondingly for $\Phi_2$. In~\eqref{eq:project}, `$\impl$'
  is sufficient, because the logic has negation. That is, we have to
  show that the set
  \begin{math}
    \{\phi\sigma_{\pi_1^-}\mid \phi\in\Phi_1\}\cup
    \{\phi\sigma_{\pi_2^-}\mid \phi\in\Phi_2\}
  \end{math}
  is one-step satisfiable. Since $\Phi_1$ and $\Phi_2$ are effectively
  finite and closed under conjunctions, it thus suffices to show that
  whenever $\phi_1\in\Phi_1$ and $\phi_2\in\Phi_2$, then
  \begin{equation*}
    \phi = \phi_1\sigma_{\pi_1^-}\land\phi_2\sigma_{\pi_2^-}
  \end{equation*}
  is one-step satisfiable. Assume the contrary; then
  $\phi_1\sigma_{\pi_1^-}\to\neg\phi_2\sigma_{\pi_2^-}$ is one-step
  valid.  Now let $\BA_1$ denote the Boolean subalgebra of $\Pow(R)$
  consisting of the $\pi_1$-invariant sets, correspondingly $\BA_2$
  for the $\pi_2$-invariant sets. One checks that $\BA_1$, $\BA_2$ are
  interpolable. Since $\Lang(\Lambda)$ has one-step interpolation, we
  therefore find $\rho\in\Prop(\Lambda(\BA_1\cap\BA_2))$ such that
  $R\models\phi_1\sigma_{\pi_1^-}\to\rho$ and
  $R\models\rho\to\neg\phi_2\sigma_{\pi_2^-}$. Using surjectivity of
  the $\pi_i$, Lemma~\ref{lem:invariance}, and compatibility, we can
  derive $\rho\sigma_{\pi_1}\in\Phi_1$, $\rho\sigma_{\pi_2}\in\Phi_2$,
  and eventually $\neg\phi_2\in\Phi_2$, contradicting satisfiability
  of~$\Phi_2$.
\end{proof}

\section{Uniform Interpolation}

We now relate one-step interpolation to interpolation for the full
logic.  Recall from Section~\ref{sec:cml} that we work in a language
with propositional variables. Given a set $V_0\subseteq V$ of
propositional variables, we write $\FLang(\Lambda,V_0)$ for the set of
$\Lambda$-formulas mentioning only propositional atoms from $V_0$, and
put
\begin{equation*}
  \FLang_n(\Lambda,V_0)=\{\phi\in\FLang(\Lambda,V_0)\mid\rank(\phi)\le
  n\}.
\end{equation*}
For a state $x$ in some model, we put
\begin{equation*}
  \Th_{V_0}^n(x)=\{\rho\in\FLang_n(\Lambda,V_0)\mid x\models\rho\}
\end{equation*}
(eliding the model, which will always be clear from the context).
Since $\Lambda$ is assumed to be finite, we have
\begin{lem}\label{lem:rank-finite}
  For finite $V_0$, $\FLang_n(\Lambda,V_0)$ is finite up to logical
  equivalence.
\end{lem}
We record explicitly:
\begin{defn}
  We say that $\Lang(\Lambda)$ has \emph{interpolation} if whenever
  $\models\phi\to\psi$ for $\phi\in\FLang(\Lambda,V_1)$ and
  $\psi\in\FLang(\Lambda,V_2)$, then there exists an \emph{interpolant}
  $\rho\in\FLang(\Lambda,V_1\cap V_2)$ such that $\models\phi\to\rho$
  and $\models\rho\to\psi$; and $\Lang(\Lambda)$ has \emph{uniform
    interpolation} if the interpolant $\rho$ can be made to depend
  only on $V_0:=V_1\cap V_2$. We then call $\rho$ a \emph{uniform
    $V_0$-interpolant} of $\phi$.
\end{defn}
We do not currently know whether one-step interpolation in the strong
sense of Definition~\ref{def:os-interpol} is necessary for
$\Lang(\Lambda)$ to have interpolation. However, a weaker version of
one-step interpolation is necessary:
\begin{lem}\label{lem:os-interpol-necessary}
  If $\Lang(\Lambda)$ has interpolation, then the one-step logic
  $\Prop(\Lambda(\Prop(V)))$ has interpolation.
\end{lem}
\noindent This can be used to disprove interpolation in some examples
(contradicting~\cite{Pattinson13} as indicated in the introduction):
\begin{expl}\label{expl:no-interpol}
  Let $\Nei_\lor$ be the subfunctor of the neighbourhood functor
  $\Nei$ defined by
  \begin{equation*}
    \Nei_\lor X=\{\alpha\in\Nei X\mid\forall A,B\subseteq X.\,A\cup B=
    X \impl (A\in\alpha\lor B\in\alpha)\},
  \end{equation*}
  and interpret the modality $\Box$ over $\Nei_\lor$ like over $\Nei.$
  Take $V_1=\{p,q\}$, $V_2=\{r,p\}$. Then the implication
  $\neg\Box(p\lor q)\to\Box(\neg p\lor r)$ is valid but has no
  interpolant in $\Prop(\{\Box\}(\Prop\{p\}))$.
\end{expl}
As to sufficiency, we have
\begin{thm}\label{thm:interpol}
  If $\Lang(\Lambda)$ has one-step interpolation then $\Lang(\Lambda)$
  has uniform interpolation. 
\end{thm}
\begin{proof}[Proof (Sketch)]
  Induction on the rank, proving the stronger claim that the rank of
  the uniform interpolant of $\phi$ is at most $\rank(\phi)$. Let
  $\phi\in\Lang_n(\Lambda,V_1)$, and let $V_0\subseteq V_1$.  We claim
  that
  \begin{equation*}
    i(\phi)=\Land\{\phi'\in\FLang_n(\Lambda,V_0)\mid\;
    \models\phi\to\phi'\}
  \end{equation*}
  (by Lemma~\ref{lem:rank-finite}, effectively a finite formula) is a
  uniform $V_0$-interpolant for $\phi$. The proof reduces
  straightforwardly to showing that, given
  $\psi\in\FLang(\Lambda,V_2)$ where $V_1\cap V_2\subseteq V_0$ and
  models $D=(Y,\zeta,\tau_2)$, $C=(X,\xi,\tau_1)$ and $y_0\in Y$,
  $x_0\in X$ such that $y_0\models_D i(\phi)\land\psi$ and
  $x_0\models_C\phi\land\Th_{V_0}(y_0)$, the formula $\phi\land\psi$
  is satisfiable.

  Using a minor variation of standard model constructions in
  coalgebraic modal
  logic~\cite{Schroder07,SchroderPattinson09,MyersEA09}, we can assume
  that $C$, $D$ are finite dags in which all states have a
  well-defined \emph{height} (distance from any initial state in a
  supporting Kripke frame), with $x_0$ and $y_0$ being initial states
  whose depth (length of the longest path starting at $x_0$ and $y_0$,
  respectively) equals the rank of the relevant formulas, and in which
  every state $x$ of height $n-k$ in $C$ is uniquely determined (among
  the states of height $n-k$) by $\Th^k_{V_1}(x)$, correspondingly for
  $y\in D$ and $\Th^k_{V_2}(x)$. Moreover, we can assume that the
  models are \emph{canonical}, i.e.\ every maximally satisfiable
  subset of $\FLang_k(\Lambda,V_1)$ is indeed satisfied at a unique
  state of height $n-k$ of $C$, as by Lemma~\ref{lem:rank-finite},
  there are only finitely many such sets; correspondingly for $D$ and
  $\FLang_k(\Lambda,V_2)$.  

  We now construct a model $E=(Z,\gamma,\tau)$ of $\phi\land\psi$ as
  follows. We put
  \begin{align*}
    Z & =\{(x,y)\in X\times Y\mid n\ge\hgt(x) =\hgt(y)=:k,
        \Th_{V_0}^{n-k}(x)=\Th_{V_0}^{n-k}(y)\}\\
      &\cup\{y\in Y\mid \hgt(y)> n\},
  \end{align*}
  denoting the first part by $Z_0$ and the second by $Z_1$, and their
  height-$k$ levels by $Z^k_0$, $Z^k_1$, $Z^k$, respectively. Note
  that $(x_0,y_0)\in Z_0$. It is straightforward to define the
  valuation $\tau$ on~$Z$. Moreover, we define a coalgebra structure
  $\gamma:Z\to\fun Z$ such that $\gamma(z)\in \fun Z^{k+1}$ for
  $z\in Z^k$. We put
  \begin{math}
    \gamma(y) = \zeta(y)\in\fun Z_1\subseteq \fun Z
  \end{math}
  for $y\in Z_1$ (using preservation of inclusions by $\fun$), and on
  states $(x,y)\in Z_0$ of maximal height $n$ by
  \begin{math}
    \gamma(x,y)=\zeta(y).
  \end{math}
  On the rest of $Z_0$ we define $\gamma$ by a \emph{coherence}
  requirement: By construction of $Z$, we have a well-defined a
  \emph{pseudo-satisfaction} relation $\models^0$ on $Z$ given by
  \begin{align*}
    (x,y) \models^0\rho & \iff
                          \begin{cases}
                            x\models_C\rho & (\rho\in\FLang_{n-\hgt(x)}(\Lambda,V_1))\\
                            y\models_D\rho & (\rho\in\FLang_{\rank(\psi)-\hgt(y)}(\Lambda,V_2))\\
                          \end{cases}\\ 
    y\models^0\rho &\iff y\models_D\rho \qquad(\rho\in\FLang_{\rank(\psi)-\hgt(y)}(\Lambda,V_2))
  \end{align*}
  For a
  $\rho\in\FLang_n(\Lambda,V_1)\cup\FLang_{\rank(\psi)}(\Lambda,V_2)$,
  we then have the \emph{pseudo-extension} $\hat\rho\subseteq Z$
  defined by
  \begin{equation*}
    \hat\rho=\{z\in Z\mid \rank(\rho)\le n-\hgt(z),
    z\models^0\rho\}.
  \end{equation*}
  Then we say that $\gamma$ is \emph{coherent} if for 
  $\hearts\rho\in\FLang_{n-k}(\Lambda,V_1)\cup\FLang_{n-k}(\Lambda,V_2)$,
  $k\le n$, and $(x,y)\in Z^k_0$,
  \begin{equation*}
    \gamma(x,y)\models\hearts(\hat\rho\cap Z^{k+1}_0)\iff(x,y)\models^0\hearts\rho.
  \end{equation*}
  For $i=0,1,2$, let $\BA_i$ be the Boolean subalgebra of
  $\Pow Z^{k+1}_0$ of sets of the form $\hat\rho\cap Z^{k+1}_0$ for
  $\rho\in\FLang_{n-k-1}(\Lambda,V_i)$. Then, of course,
  $\BA_1\cap\BA_2\subseteq\BA_0$, and by induction, $\BA_1$, $\BA_2$
  are interpolable. We define $\phi_0\in\Prop(\Lambda(\BA_1))$ and
  $\psi_0\in\Prop(\Lambda(\BA_2))$ by
  \begin{align*}
    \phi_0& =\Land\{\epsilon\hearts(\hat\rho\cap Z^{k+1}_0)\mid
            (x,y)\models^0\epsilon\hearts\rho,\rho\in\FLang_{n-k-1}(\Lambda,V_1),\epsilon\in\{\cdot,\neg\}\}\\
    \psi_0& =\Land\{\epsilon\hearts(\hat\rho\cap Z^{k+1}_0)\mid
            (x,y)\models^0\epsilon\hearts\rho,\rho\in\FLang_{n-k-1}(\Lambda,V_2),\epsilon\in\{\cdot,\neg\}\}.
  \end{align*}
  By applying Lemma~\ref{lem:os-unif-interpol} and forming the uniform $\BA_0$-interpolant $i(\phi_0)$ of $\phi_0$, and
  showing satisfiability of $i(\phi_0)\land\psi_0$, one can show that
  a coherent $\gamma$ exists. Then, we have by induction on
  $\rho\in\FLang_n(\Lambda,V_1)\cup\FLang_{\rank(\psi)}(\Lambda,V_2)$
  that $z\models_E\rho$ iff $z\models^0\rho$ for $\hgt(z)=k$ and
  $\rank(\rho)\le n-k$; so in particular $z_0=(x_0,y_0)\models\phi$
  and $z_0\models\psi$, as required.
\end{proof}
\begin{rem}
  Canonical models in the sense of the above proof sketch in fact bear
  a strong resemblance to models based on the stages of the final
  sequence of functors of the type $\Pow(V_i)\times\fun$, $i=0,1$
  (e.g.~\cite{Pattinson03}). This indicates in particular that the
  proof may eventually be made to bear a relationship, via duality,
  with Ghilardi's method of graded modal algebras~\cite{Ghilardi95}.
\end{rem}
\noindent As indicated in the introduction, our results can be summed
up as follows:
\begin{thm}\label{thm:main}
  Let $\Lambda$ be finite. Then the following properties imply each
  other in sequence:
  \begin{enumerate}
  \item\label{item:main-pb} $\Lambda$ is separating and the type functor
    $\fun$ preserves finite surjective weak pullbacks.
  \item\label{item:os-interpol} $\Lang(\Lambda)$ has one-step
    interpolation
  \item $\Lang(\Lambda)$ has uniform interpolation
  \item $\Lang(\Lambda)$ has interpolation
  \item\label{item:weak-os-interpol} The one-step logic
    $\Prop(\Lambda(\Prop(V)))$ has interpolation.
  \end{enumerate}
  Moreover, if $\Lambda$ is monotone and separating, then
  \ref{item:os-interpol}.\ implies~\ref{item:main-pb}.
\end{thm}
\noindent We note that if $\Lambda$ is finite and separating, then
$\fun$ preserves finite sets. As indicated above, we suspect but
cannot currently prove that~\ref{item:weak-os-interpol}
implies~\ref{item:os-interpol}, which would make
items~\ref{item:os-interpol}--\ref{item:weak-os-interpol}
equivalent. From Theorem~\ref{thm:main}, we obtain uniform
interpolation for the following concrete logics:
\begin{expl}
  \begin{enumerate}
  \item Whenever $\fun$ preserves weak pullbacks and $\Lambda$ is
    finite and separating, then $\Lang(\Lambda)$ has uniform
    interpolation. This case is covered already in~\cite{MartiMSc},
    see Remark~\ref{rem:lax}. In particular, we obtain that the modal
    logics $K$ and $KD$ have uniform interpolation, thus reproving
    previous results~\cite{Ghilardi95,Visser96}.
  \item Since the monotone neighbourhood functor preserves surjective
    weak pullbacks (Section~\ref{sec:pullbacks}), we obtain that
    monotone modal logic has uniform interpolation, again reproving a
    previous result~\cite{SantocanaleVenema10}.
  \item If $M$ is a finite refinable monoid, then the monoid-weighted
    functor $\Wfun{M}$ (Section~\ref{sec:pullbacks}) preserves
    surjective weak pullbacks, so that any rank-$1$ modal logic that
    is expressive (i.e.\ separating) for $\Wfun{M}$ has uniform
    interpolation, such as the logic with modalities $[m]$ for
    $m\in M$, interpreted by the predicate lifting given by
    $[m]_X(A)=\{\mu\in\Wfun{M}\mid\sum_{x\in A}\mu(x)=m\}$. This holds
    in particular when $M$ is a finite Abelian group, in which case
    $\Wfun{M}$ does not have a separating set of monotone predicate
    liftings so that this case is not covered by existing generic
    results~\cite{MartiMSc}. If we take $M=\Int/n\Int$, then the
    modalities $[m]$ described above are modulo-constraints as found
    in Presburger modal logic~\cite{DemriLugiez10}: $[m]\phi$ says
    that the number of successors of the current state (counting
    multiplicities) equals $m$ modulo $n$.
  \item Neighbourhood logic fails to preserve surjective weak
    pullbacks (Example~\ref{expl:no-surj-weak-pb}) but does have
    one-step interpolation (Example~\ref{expl:nbhd-interpol}), so we
    obtain that neighbourhood logic has uniform interpolation.
  \item One-step interpolation has been proved, in slightly different
    terms, for coalition logic~\cite{PattinsonSchroder10}, so that our
    results improve the known interpolation result for coalition
    logic~\cite{PattinsonSchroder10} to uniform interpolation.
  \end{enumerate}
\end{expl}

% \noindent Moreover, Theorem~\ref{thm:main} also yields negative
% examples via the preservation condition, such as the following.
% \begin{expl}
%   We have seen that the functor $\fun^{3}_2$ fails to preserve finite
%   surjective weak pullbacks (Example~\ref{expl:no-surj-weak-pb}). It
%   does however clearly admit a separating set of monotone modalities,
%   e.g.\ the modalities $[i]\phi$ `$\phi$ holds in the $i$-th component
%   of the successor structure', for $i=1,\dots,3$. By
%   Theorem~\ref{thm:main}, the arising logic fails to have
%   interpolation.
% \end{expl}

\begin{rem}\label{rem:lax}
  We conclude with a more detailed discussion of the relationship
  between our results and results on the logic of quasi-functorial lax
  liftings. Glossing over the ramifications of the axiomatics, a
  \emph{diagonal-preserving lax lifting} $L$ for a set functor
  $T$~\cite{MartiVenema15} extends $T$ to act also on relations,
  satisfying monotonicity w.r.t.\ inclusion of relations, preservation
  of relational converse and diagonal relations, and lax preservation
  of composition ($LR\circ LS\subseteq L(R\circ S)$). The monotone
  neighbourhood functor and its polyadic variants have
  diagonal-preserving lax liftings, and diagonal-preserving lax
  liftings are easily seen to be inherited along products and
  subfunctors, so that every functor that has a separating set of
  monotone predicate liftings has a diagonal-preserving lax
  lifting. Conversely, every finitary functor that has a
  diagonal-preserving lax lifting has a separating set of monotone
  predicate liftings, the so-called Moss liftings~\cite{MartiMSc}. A
  lax lifting induces a modal logic with a slightly non-standard
  modality $\nabla$ that generalizes Moss' modality for
  weak-pullback-preserving functors~\cite{Moss99}; for functors that
  preserve finite sets, the $\nabla$-modality and the
  predicate-lifting based modalities are however mutually
  intertranslatable~\cite{MartiMSc}, essentially by dint of the fact
  that both are separating. Summing up, for a functor that preserves
  finite sets, a diagonal-preserving lax lifting exists iff a
  separating (finite) set of monotone predicate lifting exists, and
  the induced logics are essentially the same.

  Marti~\cite{MartiMSc} shows that the logic of a diagonal-preserving
  lax lifting $L$ for $T$ has uniform interpolation if $T$ preserves
  finite sets and $L$ is \emph{quasifunctorial}, i.e.\ satisfies
  $LS\circ LR=L(S\circ R)\cap(\domain(LR)\times\range(LS))$ where
  $\domain(LR)=\{t\mid\exists t.\,(s,t)\in LR\}$ and
  $\range(LS)=\{t\mid\exists s.\,(s,t)\in LS\}$.  We recall again that
  our reduction of uniform interpolation to one-step interpolation
  holds also in cases where either separation or monotonicity fails,
  such as coalition logic / alternating-time logic and neighbourhood
  logic, respectively. Also, we have seen examples
  (Abelian-group-weighted functors) where there is no monotone
  separating set of predicate liftings but we nevertheless obtain
  uniform interpolation from preservation of surjective weak
  pullbacks.
\end{rem}
\section{Conclusions}

We have given sufficient criteria for a rank-$1$ modal logic (with
finitely many modalities), i.e.\ a coalgebraic modal logic, to have
uniform interpolation: In the general case, we have established a
reduction to the one-step logic; and in the case where the modalities
are separating, we have given a simple semantic criterion, namely
preservation of (finite) surjective weak pullbacks, which in the
monotone case is in fact also necessary for one-step interpolation. We
have thus reproved uniform interpolation for the relational modal
logics $K$ and $KD$ and for monotone (neighbourhood) modal logic, and
newly established uniform interpolation for coalition logic,
neighbourhood logic (i.e.\ classical modal logic), and various logics
of finite-monoid-weighted transition systems. All proofs are entirely
semantic; we leave a proof-theoretic treatment, in generalization of
tentative results based on cut-free sequent
systems~\cite{PattinsonSchroder10}, for future work. In particular,
such a treatment will hopefully lead to practically feasible
algorithms for the computation of interpolants. Another open question
is how our results relate to definability of bisimulation
quantifiers. \bigskip

\noindent\textbf{\sffamily Acknowledgments}\quad The authors wish to thank Tadeusz
Litak and Sebastian Enqvist for helpful discussions.
  \bibliographystyle{myabbrv} \bibliography{coalgml}

%%
%% Bibliography
%%

%% Either use bibtex (recommended), 

%\bibliography{lipics-v2016-sample-article}

%% .. or use the thebibliography environment explicitely
\newpage
\appendix
\section{Omitted Proofs}

% \paragraph*{Proof of Lemma~\ref{lem:restriction}}
%   For $\rho\in\Prop(V_0)$, we have
%   \begin{equation}
%     f^{-1}[\rho\ev_{V_0}]=\rho\ev_{V_1}. \label{eq:pres-rank0}
%   \end{equation}
%   This is shown by induction over $\rho$, with trivial Boolean cases;
%   for the atomic case, we have $\kappa\in f^{-1}[\ev_{V_0}(a)]$ iff
%   $f(\kappa)(a)=\top$ iff $\kappa(a)=\top$ iff
%   $\kappa\in\ev_{V_1}(a)$. For the main claim, we proceed by induction
%   over $\phi$, with trivial Boolean cases. For the modal case
%   $\hearts\rho$ (with $\rho\in\Prop(V_0)$), we have
%   \begin{equation*}
% t\models\hearts\rho\ev_{V_1}\stackrel{\eqref{eq:pres-rank0}}{=}
%     \hearts f^{-1}[\rho\ev_{V_0}]\iff \fun f(t)\models \hearts\rho\ev_{V_0}
%   \end{equation*}
%   where the last step is by naturality. \qed

\paragraph*{Proof of Lemma~\ref{lem:can-mod}}

We note that for $A\in\BA$,
$\can_\BA(A)=\{U\in\Ats(\BA)\mid U\subseteq A\}$.

We have a map $p:X\to\Ats(\BA)$ that maps $x\in X$ to the unique atom
$U\in\Ats(\BA)$ such that $x\in U$. We show more generally that for
the interpretations of $\phi$ and $\phi\can_\BA$,
$\phi=(\fun p)^{-1}\phi\can_\BA$. We use induction over $\phi$, with
trivial Boolean steps. For the modal case, we note that for $A\in\BA$,
$A=p^{-1}[\can_{\BA}(A)]$; the claim then follows by naturality of
predicate liftings. \qed

\paragraph*{Proof of Lemma~\ref{lem:restriction}}

We note that the canonical projection $f:\Ats(\BA_1)\to\Ats(\BA_0)$
maps $U\in\Ats(\BA_1)$ to the unique $\BA_0$-atom $V$ such that
$U\subseteq V$.

We prove the claim by induction over $\phi$, with trivial Boolean
steps. For the modal case, we note that for $A\in\BA_0$,
$\can_{\BA_1}(A)=f^{-1}[\can_{\BA_0}(A)]$. The claim follows by
naturality of predicate liftings. \qed

\paragraph*{Proof of Lemma~\ref{lem:mss}}

Define the natural transformation $\eta: \fun\to M^\Lambda_{\fun}$ by
$\eta_{X}(t)=\{\phi\in\Prop(\Lambda(\Pow(X)))\mid t\models\phi\}$,
i.e., $t\in \fun X$ is mapped to its \emph{theory} in
$M^\Lambda_{\fun}(X)$. Given a set $X$, $\eta_{X}$ is surjective by
satisfiability of sets in $M^\Lambda_{\fun}(X)$, and injective since
$\Lambda$ is separating and each $t\in\fun X$ is uniquely determined
by the set $\{\phi\in \Lambda(\Pow(X))\mid t\models \phi\}$. Hence
$\fun$ and $M^\Lambda_{\fun}$ are naturally isomorphic. \qed

\paragraph*{Proof of Lemma~\ref{lem:finitary-pb}}
  We prove only the claim for surjective pullbacks. So let 
  \begin{center}
    \begin{tikzcd}
      P \arrow{d}[left]{\pi_2}\arrow{r}{\pi_1} & X\arrow{d}{f}\\
      Y \arrow{r}[below]{g} & Z
    \end{tikzcd}
  \end{center}
  be a surjective pullback, and let $s\in \fun X$, $t\in\fun Y$ such
  that $\fun f(s)=\fun g(t)$. We have to show that there exists
  $u\in\fun Z$ such that $\fun\pi_1(u)=s$ and $\fun\pi_2(iu)=t$. Since
  $\fun$ is finitary (and we assume that $\fun$ preserves inclusions),
  there exist finite subsets $X_0\subseteq X$, $Y_0\subseteq Y$ such
  that $s\in\fun X_0\subseteq \fun X$, $t\in\fun Y_0\subseteq \fun Y$.
  Since $f$ and $g$ are surjective, we can moreover ensure, by
  suitably enlarging $X_0$ and $Y_0$, that $f[X_0]=g[Y_0]=:Z_0$. Take
  $P_0=P\cap\pi_1^{-1}[X_0]\cap\pi_2^{-1}[Y_0]$. Then
  \begin{center}
    \begin{tikzcd}
      P_0 \arrow{d}[left]{\pi_2}\arrow{r}{\pi_1} & X_0\arrow{d}{f}\\
      Y_0 \arrow{r}[below]{g} & Z_0
    \end{tikzcd}
  \end{center}
  (where we abuse $f,g,\pi_1,\pi_2$ to denote their restrictions) is a
  finite surjective pullback, hence weakly preserved by $\fun$; it
  follows that $u\in\fun P_0\subseteq\fun P$ exists as required. \qed

\paragraph*{Proof of Lemma~\ref{prop:mon}}
  Let
  \begin{center}
    \begin{tikzcd}
      P \arrow{d}[left]{\pi_2}\arrow{r}{\pi_1} & X\arrow{d}{f}\\
      Y \arrow{r}[below]{g} & Z
    \end{tikzcd}
  \end{center}
  be a surjective pullback, and let $\alpha_1\in\Mon X$,
  $\alpha_2\in\Mon Y$ such that $\Mon f(\alpha_1)=\Mon g(\alpha_2)$. We put
  \begin{equation*}
    \beta=\Up(\{\pi_{1}^{-1}[U]\mid U\in \alpha_1\}\bigcup\{\pi_{2}^{-1}[V]\mid V\in \alpha_{2}\})\in\Mon P, 
  \end{equation*}
 and claim that
  $\Mon\pi_{1}(\beta)=\alpha_{1}$ and
  $\Mon\pi_{2} (\beta)=\alpha_{2}$. By symmetry, it suffices to prove
  $\Mon\pi_{1}(\beta)=\alpha_{1}$.  Here, `$\supseteq$' is clear by
  construction of $\beta$; we prove `$\subseteq$'. We distinguish the
  following cases for $U\in \Mon\pi_{1}(\beta)$:

  If $\pi_{1}^{-1}[U']\subseteq \pi_{1}^{-1}[U]$ for some
  $U'\in \alpha_{1}$, then
  $U'=\pi_{1}\pi_{1}^{-1}[U']\subseteq \pi_{1}\pi_{1}^{-1}[U]=U$ by
  surjectivity of $\pi_1$, and hence $U\in \alpha_{1}$.

  Otherwise, $\pi_{2}^{-1}[V]\subseteq \pi_{1}^{-1}[U]$ for some
  $V\in \alpha_{2}$. Then $\pi_1[\pi_2^{-1}[V]]\in\alpha_1$ by the
  compatibility lemma, and
  $\pi_1[\pi_2^{-1}[V]]\subseteq\pi_1[\pi_1^{-1}[U]]=U$ by
  surjectivity of $\pi_1$, so $U\in\alpha_1$. \qed
  
  \paragraph*{Proof of Lemma~\ref{lem:compatible}}
  \emph{`Only if':} For $U\in\alpha_1$, we have
  $f^{-1}[f[U]]\supseteq U$ and therefore
  $f[U]\in\Mon f(\alpha_1)=\Mon g(\alpha_2)$, so
  $g^{-1}[f[U]]\in\alpha_2$; but $g^{-1}[f[U]]=\pi_2[\pi_1^{-1}[U]]$
  by the pullback property. 
  % Proof of that last bit: subset: b in lhs = > ex a in U. g(b) =
  % f(a) => ex z. pi_1 z=a, pi_2 z = b => b in rhs . 
  % supset: b in in rhs => ex z. b = pi_2 z, pi_1 z in U
  % => g b = f pi_1 z => b in lhs
  The other direction is symmetric. 

  \emph{`If':} For the inclusion
  $\Mon f(\alpha_1)\subseteq\Mon g(\alpha_2)$, let
  $U\in\Mon f(\alpha_1)$. Then $f[f^{-1}[U]]=U$ by surjectivity, and
  $f^{-1}[U]\in\alpha_1$, hence
  $g^{-1}[U]=g^{-1}[f[f^{-1}[U]]]=\pi_2[\pi_1^{-1}[f^{-1}[U]]]\in\alpha_2$
    by compatibility, so $U\in\Mon g(\alpha_2)$. The reverse inclusion
    is shown symmetrically.  \qed

\paragraph*{Proof of Lemma~\ref{lem:refinable}}

It only remains to show `if'; by Lemma~\ref{lem:finitary-pb}, it
suffices to show preservation of finite weak surjective pullbacks.

It is easy to see that refinability is exactly the condition that
ensures that $\Wfun{M}$ maps non-empty binary products to weak
pullbacks. Now $\Wfun{M}$ is generally
\emph{additive}~\cite{CoumansJacobs10}, i.e.\ maps finite coproducts
to products. Since every finite set is a coproduct of singletons,
every finite surjective pullback is a sum of non-empty finite
products. Thus, $\Wfun{M}$ maps finite surjective pullbacks to
products of weak pullbacks, which are again weak pullbacks. \qed
\paragraph*{Proof of Lemma~\ref{prop:positive}}

  \emph{`If':} If $M$ is refinable and positive, then $\Wfun{M}$
  preserves all weak pullbacks~\cite{GummSchroder01} and hence has a
  separating set of monotone predicate liftings~\cite{KurzLeal12}.

  \emph{`Only if':} Let $a+b=0$, $a\neq 0\neq b$ in $M$. We can
  understand the elements of $\Wfun{M}(X)$ as formal linear
  combinations over $X$ with coefficients from $M$. Recall that a
  monotone $n$-ary predicate lifting $\lambda$ for $\Wfun{M}$ is
  equivalently determined by a subset of
  $\Wfun{M}(2^n)$ that is
  closed under the relation $\lesssim$ given by taking $t\lesssim s$
  if $t,s$ can be written as linear combinations
  $s=\sum_{i=1}^na_ix_i$, $t=\sum_{i=1}^na_iy_i$ such that
  $x_i\le y_i$ for $i=1,\dots,n$ in the pointwise ordering
  on~$2^n$~\cite{Schroder08}. Now let $X$ be a set with at least two
  distinct elements $x\neq y$. We show that $a x,a y\in\Wfun{M}X$ are
  indistinguishable under $n$-ary monotone predicate liftings by
  showing that for any $f:X\to 2^n$, we have
  $Tf(a x)\lesssim Tf(a y)\lesssim Tf(a x)$. Namely: Writing
  $\bot^n\in 2^n$ for the $n$-tuple $(\bot,\dots,\bot)$, we have
  \begin{equation*}
    Tf(a x)  = af(x) = af(x)+a\bot^n+b\bot^n \lesssim af(x)+af(y)+bf(x)
    = af(y)=Tf(a y),
  \end{equation*}
  and analogously $Tf(ay)\lesssim Tf(ax)$. \qed

\paragraph*{Proof of Lemma~\ref{lem:os-unif-interpol}}
  `Only if' is trivial; we prove `if'. So let
  $\phi\in\Prop(\Lambda(\Prop(V_1)))$, and let $V_0\subseteq V_1$. We
  put
  \begin{equation*}
    i(\phi)=\Land\{\rho\in\Prop(\Lambda(\Prop(V_0)))\mid\models\phi\to\rho\},
  \end{equation*}
  effectively a finite formula because $\Lambda$ is finite and $\Prop$
  preserves finite sets up to logical equivalence, and claim that
  $i(\phi)$ is a uniform $V_0$-interpolant of $\phi$. We clearly have
  $\models\phi\to i(\phi)$. Now let
  $\psi\in\Prop(\Lambda(\Prop(V_2)))$ where
  $V_1\cap V_2\subseteq V_0$, and let $\models\phi\to\psi$. By
  one-step interpolation, we have $\rho\in\Prop(V_0)$ such that
  $\models\phi\to\rho$ and $\models\rho\to\psi$. But then $\rho$ is a
  conjunct of $i(\phi)$, so $\models i(\phi)\to\psi$, as required.  \qed

\paragraph*{Proof of Lemma~\ref{lem:interpol-implies-preservation}}

  Let 
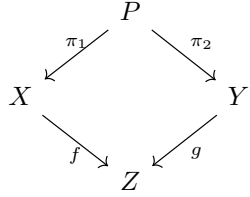
  \begin{center}
    \begin{tikzcd}
      & P \arrow{dl}[above]{\pi_1}\arrow{dr}{\pi_2}\\
      X \arrow{dr}[below]{f} & & Y\arrow{dl}{g}\\
      & Z
    \end{tikzcd}
  \end{center}
  be a finite surjective
  pullback. % ; we understand $P$ as a difunctional
  % relation between $X$ and $Y$, in particular we assume
  % $P\subseteq X\times Y$
  As indicated in Section~\ref{sec:prelims}, we can identify $\fun$ with
  its MSS functor, i.e.\ we assume that $\fun X$ consists of maximally
  satisfiable subsets
  $\Phi\subseteq\Prop(\Lambda(\Pow(X)))$. % Of course, $\Phi$ is
  % completely determined by its restriction to $\Lambda(\Pow(X))$. We
  % assume moreover that $\Lambda$ is closed under duals, and work with
  % negation normal forms when convenient.

  In this reading, we are given $\Phi_1\in \fun X$ and
  $\Phi_2\in \fun Y$ such that $\fun f(\Phi_1)=\fun g(\Phi_2)$, which
  by a straightforward generalization of Lemma~\ref{lem:compatible}
  means that $\Phi_1$ and $\Phi_2$ are \emph{compatible}, i.e.\
  \begin{equation*}
    \phi\in\Phi_2\text{ implies }\phi\sigma_{\pi_2^-}\sigma_{\pi_1}\in\Phi_1
  \end{equation*}
  and symmetrically. We have to show that there exists
  $\Phi\in \fun P$ such that $F\pi_1(\Phi)=\Phi_1$ and
  $F\pi_2(\Phi)=\Phi_2$, i.e.\
  \begin{equation*}
    \phi\in\Phi_1\iff \phi\sigma_{\pi_1^-}\in\Phi
  \end{equation*}
  and correspondingly for $\Phi_2$. Observe that in the above
  condition, `$\implies$' is sufficient, because the logic has
  negation.

  That is, we have to show that the set
  \begin{equation*}
    \{\phi\sigma_{\pi_1^-}\mid \phi\in\Phi_1\}\cup
    \{\phi\sigma_{\pi_2^-}\mid \phi\in\Phi_2\}
  \end{equation*}
  is one-step satisfiable. Since $\Phi_1$ and $\Phi_2$ are effectively
  finite and closed under conjunctions, it thus suffices to show that
  whenever $\phi_1\in\Phi_1$ and $\phi_2\in\Phi_2$, then
  \begin{equation*}
    \phi = \phi_1\sigma_{\pi_1^-}\land\phi_2\sigma_{\pi_2^-}
  \end{equation*}
  is one-step satisfiable. Assume the contrary; then
  $\phi_1\sigma_{\pi_1^-}\to\neg\phi_2\sigma_{\pi_2^-}$ is one-step
  valid.  Now let $\BA_1$ denote the Boolean subalgebra of $\Pow(P)$
  consisting of the $\pi_1$-invariant sets, correspondingly $\BA_2$
  for the $\pi_2$-invariant sets.  Since $\Lang(\Lambda)$ has one-step
  interpolation, we therefore find
  $\rho\in\Prop(\Lambda(\BA_1\cap\BA_2))$ such that
  $P\models\phi_1\sigma_{\pi_1^-}\to\rho$ and
  $P\models\rho\to\neg\phi_2\sigma_{\pi_2^-}$.  Now by
  Lemma~\ref{lem:invariance},
  \begin{equation*}
    X\models\phi_1\sigma_{\pi_1^-}\sigma_{\pi_1}\to\rho\sigma_{\pi_1}.
  \end{equation*}
  Since $\pi_1$ is surjective,
  $\phi_1\sigma_{\pi_1^-}\sigma_{\pi_1}=\phi_1$, and since
  $\phi_1\in\Phi_1$, it follows by maximal satisfiability that
  $\rho\sigma_{\pi_1}\in\Phi_1$, so by compatibility,
  \begin{equation}\label{eq:rho-in-phi2-app}
    \rho\sigma_{\pi_1}\sigma_{\pi_1^-}\sigma_{\pi_2}\in\Phi_2.
  \end{equation}
  We claim that for $A\in\BA_1$,
  \begin{equation*}
    (\sigma_{\pi_1}\sigma_{\pi_1^-}\sigma_{\pi_2})(A)=\sigma_{\pi_2}(A).
  \end{equation*}
  Indeed, keeping in mind that composition of substitutions is
  diagrammatic, we have 
  \begin{equation*}
    \pi_2[\pi_1^{-1}[\pi_1[A]]]=\pi_2[A]
  \end{equation*}
  by $\pi_1$-invariance of $A$. Thus, from~\eqref{eq:rho-in-phi2-app}, we
  obtain $\rho\sigma_{\pi_2}\in\Phi_2$. Since
  $\rho\to\neg\phi_2\sigma_{\pi_2^-}\in\Prop(\Lambda(\BA_2))$, we have
  by Lemma~\ref{lem:invariance}
  \begin{equation*}
    X\models\rho\sigma_{\pi_2}\to\neg\phi_2\sigma_{\pi_2^-}\sigma_{\pi_2},
  \end{equation*}
  and since, again, $\phi_2\sigma_{\pi_2^-}\sigma_{\pi_2}=\phi_2$
  because $\pi_2$ is surjective, it follows that
  $\neg\phi_2\in\Phi_2$, in contradiction to satisfiability of
  $\Phi_2$. \qed
\paragraph*{Proof of Lemma~\ref{lem:invariance}}

  Induction over $\phi$, with trivial Boolean steps. For the modal
  step, i.e.\ by the above for atoms of the form $\hearts f^{-1}[B]$,
  note that $\fun f$ is also surjective, and for $t\in \fun X$,
  \begin{equation*}
    s\models \hearts f^{-1}[B]\iff \fun f(s)\models\hearts B =\hearts f[f^{-1}[B]], 
  \end{equation*}
  where the last equality is by surjectivity. \qed

\paragraph*{Proof of Lemma~\ref{lem:os-interpol-necessary}}
  Let $\phi\in\Prop(\Lambda(\Prop(V_1)))$ and
  $\psi\in\Prop(\Lambda(\Prop(V_2)))$ such that
  $\models\phi\to\psi$. Since $\Lang(\Lambda)$ has propositional
  variables, we can see $\phi$ and $\psi$ as formulas in
  $\FLang(\Lambda)$, so $\phi$ and $\psi$ have an interpolant
  $\theta\in\FLang(\Lambda,V_1\cap V_2)$

  We proceed to show that we can modify $\theta$ to obtain an
  interpolant in $\Prop(\Lambda(\Prop(V_1\cap V_2)))$. We begin by
  eliminating occurrences of propositional variables at the top level
  of~$\theta$, i.e.\ outside the scope of modal operators. Let
  $\theta'$ arise from $\theta$ by replacing all such occurrences with
  $\top$. Then, of course, $\theta'\in\FLang(\Lambda,V_1\cap V_2)$.
  Moreover, we still have $\models\phi\to\theta'$: Let $x$ be a state
  in a coalgebraic model~$C$ such that $C,x\models\phi$; we have to
  show $C,x\models\theta'$. Since coalgebraic modal logic has the tree
  model property~\cite{SchroderPattinson09a}, we can assume that $C$ is
  a tree. Under this assumption, we can change the valuation of all
  propositional variables at $x$ to~$\top$ without affecting
  satisfaction of~$\phi$, obtaining $C'$ such that
  $C',x\models\phi$. Then $C',x\models\theta$, which by the choice of
  valuation at $x$ implies $C',x\models\theta'$. Since $\theta'$ does
  not contain propositional variables at the top level and $C$ is
  tree-shaped, it follows that $C,x\models\theta'$. The proof that
  $\models\theta'\to\psi$ is entirely analogous.

  We can thus assume that
  $\theta\in\Prop(\Lambda(\FLang(\Lambda)))$. That is, $\theta$ has
  the form
  \begin{equation*}
    \theta = \theta_0\sigma
  \end{equation*}
  where 
  \begin{equation*}
    \theta_0\in\Prop(\Lambda(\Prop(V_0\cup(V_1\cap V_2))))
  \end{equation*}
  where $V_0$ is disjoint from $V_1$ and $V_2$, and
  \begin{equation*}
    \sigma:V_0\to\Lambda(\FLang(\Lambda)).
  \end{equation*}
  Informally speaking, $V_0$ and $\sigma$ represent the parts
  of~$\theta$ that stick out of rank~$1$. Now let $y$ be a state in
  some coalgebraic model $D$, and define $\sigma':V_0\to 2$ by
  \begin{equation*}
    \sigma'(v)=\top\quad\text{iff}\quad
    D,x\models\sigma(v).
  \end{equation*}
  Then $\theta':=\theta_0\sigma'$ is rank~$1$ and mentions only
  variables in $V_1\cap V_2$; we claim that it interpolates between
  $\phi$ and $\psi$.

  Again, we prove only that $\models\phi\to\theta'$. So let $x$ be a
  state in a coalgebraic model $C$ such that $C,x\models\phi$; we have
  to show $C,x\models\theta'$. As above, we can assume that $C$ is
  tree-shaped. We now construct a coalgebraic model $C'$ by replacing
  every successor state of $x$ and the associated subtree (according
  to the supporting Kripke frame of $C$) with a copy of $y$ and $D$,
  keeping however the original valuation for the propositional
  variables. Then 
  \begin{equation}\label{eq:C-prime}
    \text{$C$ and $C'$ satisfy the same rank-$1$ formulas,}
  \end{equation}
  in particular $C',x\models\phi$ and therefore
  $C',x\models\theta$. By the definition of
  $\sigma'$ it follows that $C',x\models\theta'$, and since
  $\theta'$ is rank $1$, this implies $C,x\models\theta'$, again
  by~\eqref{eq:C-prime}. \qed

\paragraph*{Details for Example~\ref{expl:no-interpol}}

We use liberally that the rule $a\lor b/\Box a\lor\Box b$ is
\emph{one-step cutfree complete} for the logic of
$\Nei_\lor$~\cite{SchroderPattinson09}, i.e.\ every valid clause over
$\{\Box\}(\Prop(V))$ must contain $\Box\phi$ and $\Box\psi$ such that
$\phi\lor\psi$ is valid. 

Assume that an interpolant in $\Prop(\{\Box\}(\Prop(\{p\}))$
exists. Then by the same argument as in
Lemma~\ref{lem:os-unif-interpol}, the formula $\chi$ obtained as the
essentially finite conjunction of all clauses $\rho$ over
$\{\Box\}(\Prop(\{p\}))$ such that $\Box(p\lor q)\lor\rho$ is
(one-step) valid is also an interpolant. In the definition of $\chi$,
we can clearly omit clauses $\rho$ that are themselves valid. Then the
only way that $\Box(p\lor q)\lor\rho$ is valid is that $\rho$ contains
(and then w.l.o.g.\ equals) a literal $\Box(\theta)$ with
$\theta\in\Prop(\{p\})$ such that $(p\lor q)\lor\theta$ is valid. Up
to logical equivalence, $\theta$ is one of $\top,\bot,p,\neg p$;
validity of $(p\lor q)\lor\theta$ leaves only $\theta=\neg p$. But
$\Box(\neg p)\to\Box(\neg p\lor r)$ is not valid. \qed

\paragraph*{Proof of Theorem~\ref{thm:interpol}}

  Let $\phi\in\Lang_n(\Lambda,V_1)$, and let $V_0\subseteq V_1$.  We
  claim that
  \begin{equation*}
    i(\phi)=\Land\{\phi'\in\FLang_n(\Lambda,V_0)\mid\;
    \models\phi\to\phi'\}
  \end{equation*}
  (by Lemma~\ref{lem:rank-finite}, effectively a finite formula) is a
  uniform $V_0$-interpolant for $\phi$.  Clearly,
  $\models\phi\to i(\phi)$, so it remains to show that
  $\models i(\phi)\to\psi$ given $\psi\in\FLang(\Lambda,V_2)$ such
  that $\models\phi\to\psi$ and $V_1\cap V_2\subseteq V_0$. We can
  clearly assume that $i(\phi)$ is satisfiable; since
  $\bot\in\FLang_n(\Lambda,V_0)$, this implies that $\bot$ is not a
  conjunct of $i(\phi)$, i.e.\ $\phi$ is satisfiable. We can then
  dualize the goal and prove that whenever $i(\phi)\land\psi$ is
  satisfiable for some $\psi\in\FLang(\Lambda,V_2)$ where
  $V_1\cap V_2\subseteq V_0$, then $\phi\land\psi$ is satisfiable. So
  we are given two models, $D=(Y,\zeta,\theta_2)$ such that
  $y_0\models_D i(\phi)\land\psi$ for some $y_0\in Y$, and
  $C=(X,\xi,\theta_1)$ such that $x_0\models_C\phi$ for some
  $x_0\in X$. Let $\rho=\Th_{V_0}(y_0)$. Then $\phi\land\rho$ is
  satisfiable (otherwise, $\models\phi\to\neg\rho$, so
  $i(\phi)\to\neg\rho$ and hence $y_0\models\neg\rho$, contradiction),
  so we assume that $x_0$ in fact satisfies $\phi\land\rho$.

  Using a minor variation of standard model constructions in
  coalgebraic modal
  logic~\cite{Schroder07,SchroderPattinson09,MyersEA09}, we can assume
  that both these models are finite dags in which all states have a
  well-defined \emph{height} (distance from any initial state in a
  smallest supporting Kripke frame, which exists because the models
  are finite), with $x_0$ and $y_0$ being initial states whose depth
  (length of the longest path starting at $x_0$ and $y_0$,
  respectively) equals the rank of the relevant formulas, and in which
  every state $x$ of height $n-k$ in $C$ is uniquely determined (among
  the states of height $n-k$) by $\Th^k_{V_1}(x)$, correspondingly for
  $y\in D$ and $\Th^k_{V_2}(x)$. Moreover, we can assume that the
  models are \emph{canonical}, i.e.\ every maximally satisfiable
  subset of $\FLang_k(\Lambda,V_1)$ is indeed satisfied at a unique
  state of height $n-k$ of $C$, as by Lemma~\ref{lem:rank-finite},
  there are only finitely many such sets; correspondingly for $D$ and
  $\FLang_k(\Lambda,V_2)$.  

  We now construct a model $E=(Z,\gamma,\theta)$ of $\phi\land\psi$ as
  follows. We put
  \begin{align*}
    Z & =\{(x,y)\in X\times Y\mid n\ge\hgt(x) =\hgt(y)=:k,
        \Th_{V_0}^{n-k}(x)=\Th_{V_0}^{n-k}(y)\}\\
      &\cup\{y\in Y\mid \hgt(y)> n\},
  \end{align*}
  denoting the first part by $Z_0$ and the second by $Z_1$, and their
  height-$k$ levels by $Z^k_0$, $Z^k_1$, $Z^k$, respectively.  Note
  that $(x_0,y_0)\in Z_0$.

  By construction, states $z\in Z$ have a well-defined height
  $\hgt(z)$. We define a $V_1\cup V_2$-valuation $\theta$ on $Z$ by
  \begin{align*}
    Z_0\owns(x,y)\in \theta(p) & \iff
                                 \begin{cases}
                                   x\in \theta_1(p) & (p\in V_1) \\
                                   y\in \theta_2(p) & (p\in V_2)
                                 \end{cases}\\
    Z_1\owns y\in \theta(p) &\iff y\in\theta_2(p),
  \end{align*}
  which is well-defined by construction of $Z_0$. Moreover, we define
  a coalgebra structure $\gamma:Z\to\fun Z$ such that
  $\gamma(z)\in \fun Z^{k+1}$ for $z\in Z^k$. We put
  \begin{equation*}
    \gamma(y) = \zeta(y)\in\fun Z_1\subseteq \fun Z
  \end{equation*}
  for $y\in Z_1$ (using preservation of inclusions by $\fun$), and on
  states $(x,y)\in Z_0$ of maximal height $n$ by
  \begin{equation*}
    \gamma(x,y)=\zeta(y).
  \end{equation*}
  On the rest of $Z_0$ we define $\gamma$ by a \emph{coherence}
  requirement: we define a \emph{pseudo-satisfaction} relation
  $\models^0$ between
  \begin{itemize}
  \item states in $Z$ of height $k$ and $\FLang(\Lambda,V_1)$-formulas
    of rank at most $n-k$, as well as
  \item states in $Z$ of height $k$ and $\FLang(\Lambda,V_2)$-formulas
    of rank at most $\rank(\psi)-k$.
  \end{itemize}
  by
  \begin{align*}
    (x,y) \models^0\rho & \iff
                          \begin{cases}
                            x\models_C\rho & (\rho\in\FLang(\Lambda,V_1))\\
                            y\models_D\rho & (\rho\in\FLang(\Lambda,V_2))\\
                          \end{cases}\\
    y\models^0\rho &\iff y\models_D\rho.
  \end{align*}
  This is well-defined by construction of $Z_0$.  For a formula
  $\rho\in\FLang_n(\Lambda,V_1)\cup\FLang_{\rank(\psi)}(\Lambda,V_2)$,
  we then have the \emph{pseudo-extension} $\hat\rho\subseteq Z$
  defined by
  \begin{equation*}
    \hat\rho=\{z\in Z\mid \rank(\rho)\le n-\hgt(z),
    z\models^0\rho\}.
  \end{equation*}
  Then we say that $\gamma$ is \emph{coherent} if for 
  $\hearts\rho\in\FLang_{n-k}(\Lambda,V_1)\cup\FLang_{n-k}(\Lambda,V_2)$,
  $k\le n$, and $(x,y)\in Z^k_0$,
  \begin{equation*}
    \gamma(x,y)\models\hearts(\hat\rho\cap Z^{k+1}_0)\iff(x,y)\models^0\hearts\rho.
  \end{equation*}
  For $i=0,1,2$, let $\BA_i$ be the Boolean subalgebra of
  $\Pow Z^{k+1}_0$ of sets of the form $\hat\rho\cap Z^{k+1}_0$ for
  $\rho\in\FLang_{n-k-1}(\Lambda,V_i)$. Then, of course,
  $\BA_1\cap\BA_2\subseteq\BA_0$, and by induction, $\BA_1$, $\BA_2$
  are interpolable. 

  We take $W$ to be the set of variables $a_\rho$ where $\rho$ ranges
  over $\FLang_{n-k-1}(\Lambda,V_1\cup V_2)$, and
  $W_i=\{a_\rho\mid \rho\in\FLang_{n-k-1}(\Lambda,V_i)\}$ for
  $i=0,1,2$. We define $\phi_0\in\Prop(\Lambda(W_1))$ and
  $\psi_0\in\Prop(\Lambda(W_2))$ by
  \begin{align*}
    \phi_0& =\Land\{\epsilon\hearts a_\rho\mid
            (x,y)\models^0\epsilon\hearts\rho,\rho\in\FLang_{n-k-1}(\Lambda,V_1),\epsilon\in\{\cdot,\neg\}\}\\
    \psi_0& =\Land\{\epsilon\hearts a_\rho\mid
            (x,y)\models^0\epsilon\hearts\rho,\rho\in\FLang_{n-k-1}(\Lambda,V_2),\epsilon\in\{\cdot,\neg\}\}.
  \end{align*}
  We let $\tau$ denote the $\Pow(Z^{k+1}_0)$-valuation given by
  \begin{equation*}
    \tau(a_\rho)=\hat\rho\cap Z^{k+1}_0.
  \end{equation*}
  It suffices to show that $(\phi_0\land\psi_0)\tau$ is
  $\tau$-satisfiable. Now $\rho\in\Prop(\Lambda(\BA_0)))$ be the
  uniform one-step $\BA_0$-interpolant of
  $\phi_0\tau\in\Prop(\Lambda(\BA_1))$, and let
  $i(\phi)\in\Prop(\Lambda(W_0))$ such that $i(\phi)\tau=\rho$. Then
  it suffices to show that $(i(\phi_0)\land\psi_0)\tau$ is
  satisfiable.
  % Detailed argument for this:
  % If phi0 & psi0 is unsat, then phi0 -> neg psi0, so 
  % phi0 tau -> neg psi0 tau, so
  Now let
  $Y^k\subseteq Y$ be the set of states of height $k$ in $D$, and let
  $\tau_1$ denote the $\Pow(Y^{k+1})$-valuation given by
  \begin{equation*}
    \tau_1(a_\rho)=\Sem{\rho}_D\cap Y^{k+1}.
  \end{equation*}
  Then
  \begin{equation*}
    \tau(a_\rho)=\pi_2^{-1}[\tau_1(a_\rho)]\qquad
    \text{for $a_\rho\in W_2$}
  \end{equation*}
  by the definition of $\hat\rho$, where $\pi_2:Z_0\to Y$ is the
  second projection, so since
  $i(\phi_0)\land\psi_0\in\Prop(\Lambda(\Prop(W_2)))$ it suffices to
  show that $(i(\phi_0)\land\psi_0)\tau_1$ is satisfiable. In fact we
  claim that $\zeta(y)\models(i(\phi_0)\land\psi_0)\tau_1$. Since for
  $\rho\in\FLang_{n-k-1}(\Lambda,V_2)$,
  $(x,y)\models^0\epsilon\hearts\rho$ iff
  $y\models\epsilon\hearts\rho$ iff
  $\zeta(y)\models\epsilon\hearts a_\rho\tau_1$, we have
  $\zeta(y)\models\psi_0\tau_1$ by definition of $\psi_0$. It remains
  to show that $\zeta(y)\models i(\phi_0)\tau_1$. Let $\sigma$ be the
  substitution defined by $\sigma(a_\rho)=\rho$. Then we have
  \begin{align*}
    & \zeta(y)\models i(\phi_0)\tau_1\\
    & \iff y\models i(\phi_0)\sigma &&\by{semantics}\\
    & \iff x\models i(\phi_0)\sigma
    && \by{$i(\phi_0)\sigma\in\FLang_{n-(k-1)}(\Lambda,V_0)$}\\
    & \Leftarrow x\models\phi_0\sigma,
  \end{align*}
  and the last condition holds by construction of $\phi_0$ because for
  $\rho\in\FLang_{n-k}(\Lambda,V_0)$,
  $(x,y)\models\epsilon\hearts\rho$ iff $x\models\epsilon\hearts\rho$. 
    This shows that a coherent $\gamma$ exists. Then, we have by
  induction on
  $\rho\in\FLang_n(\Lambda,V_1)\cup\FLang_{\rank(\psi)}(\Lambda,V_2)$
  that $z\models_E\rho$ iff $z\models^0\rho$ for $\hgt(z)=k$ and
  $\rank(\rho)\le n-k$; so in particular $z_0=(x_0,y_0)\models\phi$
  and $z_0\models\psi$, as required. \qed

\end{document}